\documentclass[number,final,3p]{elsarticle}
\usepackage[switch]{lineno}
\usepackage{relsize}
\usepackage{color}
\usepackage{algpseudocode}
\usepackage{graphicx}
\usepackage{subcaption}
\usepackage{algorithm}
\usepackage{bm}
\usepackage[colorlinks]{hyperref}% Add hyper-ref for equations, figures, etc.
\makeatletter
\gdef\urlauthor#1#2{\g@addto@macro\@elsuads{\let\corref\@gobble%
     \def\@@tmp{#1}\raggedright\eadsep
     {\ttfamily\url{\expandafter\strip@prefix\meaning\@@tmp}}\space(#2)%
     \def\eadsep{\unskip,\space}}%
}
\gdef\emailauthor#1#2{\stepcounter{ead}%
     \g@addto@macro\@elseads{\raggedright%
      \let\corref\@gobble\def\@@tmp{#1}%
      \eadsep{\ttfamily\href{mailto:\expandafter\strip@prefix\meaning\@@tmp}{\expandafter\strip@prefix\meaning\@@tmp}}
      (#2)\def\eadsep{\unskip,\space}}%
}
\makeatother
\makeatletter
\let\@afterindenttrue\@afterindentfalse
\makeatother
\usepackage{amssymb}
\usepackage{amsthm}
\usepackage{amsmath}
\usepackage{amssymb}
\usepackage{mathtools}
\usepackage{dsfont}
\usepackage{booktabs}
\usepackage{url}
\usepackage{scrextend}
\usepackage{epstopdf}
\usepackage{float}
\usepackage{tcolorbox}
\usepackage{tablefootnote}
\usepackage[perpage]{footmisc}% Renew the footnote numbering for each page
\usepackage{bbm}
\usepackage{lineno}
\usepackage{svg}
\usepackage{soul}
\def\r{\mathbb{R}}
\def\rn{\mathbb{R}^n}
\def\defi{\coloneqq}

\newcommand{\E}[1]{\mathbb{E}\left[#1\right]}

\newcommand{\vect}[1]{\boldsymbol{#1}}


\biboptions{numbers,sort&compress,square}
\journal{arXiv}
\begin{document}
% \linenumbers
	\begin{frontmatter}
		\renewcommand{\thefootnote}{\fnsymbol{footnote}}
		\title{Risk Twin: Real-time Risk Visualization and Control \\for Structural Systems}
		\author[1]{Zeyu Wang}
		\author[2]{Ziqi Wang\corref{cor1}}
         \ead{ziqiwang@berkeley.edu}
         \cortext[cor1]{Corresponding author}
        \address[1]{School of Engineering and Technology, China University of Geosciences, Beijing, China}
		\address[2]{Department of Civil and Environmental Engineering, University of California, Berkeley, United States}
		\begin{abstract}
Digital twinning in structural engineering is a rapidly evolving technology that aims to eliminate the gap between physical systems and their digital models through real-time sensing, visualization, and control techniques. Although Digital Twins can offer dynamic insights into physical systems, their accuracy is inevitably compromised by uncertainties in sensing, modeling, simulation, and control. This paper proposes a specialized Digital Twin formulation, named Risk Twin, designed for real-time risk visualization and risk-informed control of structural systems. Integrating structural reliability and Bayesian inference methods with Digital Twinning techniques, Risk Twin can analyze and visualize the reliability indices for structural components in real-time. To facilitate real-time inference and reliability updating, a \textit{simulation-free} scheme is proposed. This scheme leverages precomputed quantities prepared during an offline phase for rapid inference in the online phase. Proof-of-concept numerical and real-world Risk Twins are constructed to showcase the proposed concepts.
    \end{abstract}
	\end{frontmatter}
	
	\renewcommand{\thefootnote}{\fnsymbol{footnote}}
	
	%% main text
\section{Introduction}
	Digital twin technology involves creating a virtual replica of a physical system to simulate, predict, and understand the behavior of the modeled system, facilitating decision-making under evolving conditions \cite{grieves2014digital,grieves2017digital,tao2018digital,van2023digital}. This technology yields an increasing number of applications, such as civil infrastructures \cite{boje2020towards,lu2019digital,mohammadi2021thinking}, mechanical \cite{moi2020digital,lai2021designing} and battery \cite{thelen2022comprehensive,thelen2023comprehensive} systems, medical and health \cite{niederer2021scaling,hernandez2021digital,venkatesh2022health}, signal processing \cite{maksymenko2023myoelectric}, manufacturing \cite{xia2021digital,vatankhah2021digital}, and machine learning \cite{maksymenko2023myoelectric,san2023decentralized}. In structure and infrastructural systems, a highly relevant technology is structural health monitoring \cite{spencer2019advances,catbas2008structural}, which focuses on identifying structural damage and deterioration over time. The boundary between Digital Twining and advanced structural health monitoring systems can be blurred, but in general, applying Digital Twining technology to structural engineering projects can yield interactive cyber-physical systems with broader applications than structural health monitoring. Ideally, the interactions between physical systems and their Digital Twins are bi-directional:  sensor data collected from the physical system can lead to an updated digital model, and adjustments to control parameters in the digital model can also affect the physical system. For instance, Kapteyn et al. demonstrated the use of a Digital Twin to monitor and control the state of an unmanned aerial vehicle in real-time \cite{kapteyn2021probabilistic}. Torzoni et al. adapted the technique to civil structures, facilitating proactive decision-making regarding maintenance and management actions \cite{torzoni2024digital}. Furthermore, the interaction between physical and cyber components of a composite structure was successfully achieved by Xu et al. \cite{xu2023deep}. These studies highlight the dynamic interplay between digital and physical components, extending beyond simply representing the physical system in a digital format.
 
The inspiration for this study stems from the work of Haag and Anderl \cite{haag2018digital}, where they demonstrated a pilot cyber-physical Digital Twin system of a bending beam using real-time sensing and three-dimensional computer-aided design visualization techniques. Their system accurately displayed the bending and deformation of the beam in real time, achieving a tight linkage between the physical and Digital Twins. However, a prevailing assumption in many Digital Twin systems is that the sensing, modeling, and simulation modules are deterministic \cite{khayyati2021lab,salvi2022cyber,rodriguez2023updating,lai2023digital}. This assumption, however, is not realistic in real-world Digital Twin applications, where uncertainties are pervasive \cite{der2009aleatory}. These uncertainties arise from various sources, including measurement uncertainties due to inaccuracies and variabilities of sensors; modeling uncertainties stemming from limitations and assumptions in computational models; and operational uncertainties, which involve variations in system performance under different real-world conditions and human interactions \cite{straub2010bayesianm,straub2015bayesian}. For instance, stress-sensing devices in structural Digital Twin systems often exhibit high sensitivity to temperature and humidity, causing significant discrepancies in Analog-to-Digital (AD) conversion values for identical stress levels measured under differing conditions, such as noon versus night \cite{enander1984performance}. These uncertainties can compromise the accuracy of deterministic Digital Twin models. To properly quantify the influence of uncertainties in Digital Twin systems, Kapteyn et al. proposed a comprehensive framework based on Bayesian network \cite{kapteyn2021probabilistic}. This framework represents the states of physical systems and their digital counterparts as random variables and updates the latter through Bayesian network inference using the observed data. Moreover, two realistic applications of predictive Digital Twin are proposed to illustrate the framework, which have been further explored in subsequent research \cite{niederer2021scaling,torzoni2024digital}. 

Although the previously discussed methodologies can predict specific quantities of interest, there is an absence of a specialized framework for assessing the risk of structural safety on a unified scale. Such a framework is essential not only for quantifying uncertainties and risks within the structural system, but also for providing a holistic metric of risk across different structural components. For instance, a Digital Twin can monitor quantities such as stress, strain, and moment that collectively contribute to the safety of structural components, but it does not offer a directly comparable metric for the risk of each component. To fill this gap, this study introduces Risk Twin, a system designed to analyze and visualize the reliability indices for structural components in real-time, thereby facilitating risk-informed control.

Building on existing progress of integrating probabilistic analysis within the Digital Twinning technology, this study aims to further advance their application in risk assessment and control of civil structures. Specifically, this paper proposes (i) the concept of Risk Shadow as a digital reflection of structural risk; (ii) an efficient \textit{simulation-free} method to update the Risk Shadow; and (iii) real-world and numerical benchmark Risk Twin systems. In the benchmark experiments, the bidirectional interaction between the physical and digital systems is demonstrated. 

The paper is structured as follows: Section \ref{sec:rt} introduces the basic concepts of Risk Twin. Section \ref{sec:be} presents implementation details for two benchmark Risk Twin systems. Moreover, simulation of wind turbine RT is elaborated in Section \ref{sec:wt}. Additionally, demonstration videos for these Risk Twin systems are provided for further reference. Consequently, conclusive remarks are drawn in Section \ref{Sec:conclude}.

\section{Risk Twin}\label{sec:rt}
If a Digital Twin operates with deterministic parameters, it will classify the state of a system as either \textit{safe} or \textit{failure}. However, uncertainties are prevalent in Digital Twin systems due to variabilities in sensing and modeling, hindering an exact reflection of the true state of a system. Risk Twin is a probabilistic Digital Twin designed to provide real-time quantification, visualization, and control of risks for the modeled system. This section begins with a general introduction to the framework of Risk Twin and then delves into the detailed developments of its core components.

\subsection{General framework}
\begin{figure}[t]
  \centering
  \includegraphics[scale=0.8]{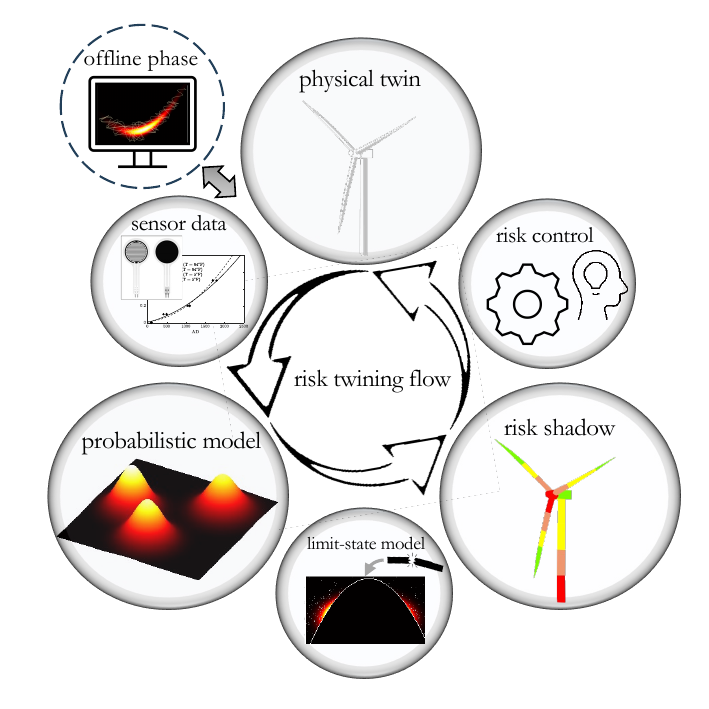}
  \caption{\textbf{Workflow of the Risk Twin}. The workflow begins with sensors collecting data from the physical twin. This data is then used to update the probabilistic model of basic random variables. The updated probabilistic model is propagated into a reliability analysis module to obtain reliability indices for system components. Subsequently, the Risk Shadow, displaying these reliability indices, is presented to users. Users can then decide whether to execute control operations upon perceiving the Risk Shadow. Once a control command is received, actuators, governed by an algorithm with operational cost constraints, act on the physical system, completing the full loop of information flows between the physical and digital systems. To address the computational challenge of real-time inference and reliability updating, the operation of Risk Twin is decomposed into offline and online phases. During the offline phase, computationally intensive models are simulated to prepare datasets and quantities for rapid, simulation-free inference in the online phase.}
  \label{Fig:1}
\end{figure}

Risk Twin, illustrated in Figure \ref{Fig:1}, encompasses the forward information flow from the physical system to the digital model and the inverse flow from the digital model to the physical system. The forward flow includes (i) Bayesian inference – statistical inference for basic random variables, and (ii) Risk Shadow – the digital representation of risk. The inverse flow involves the risk-informed control of the physical system. The computational challenges lie in the efficient statistical inference for basic random variables and reliability indices. Specialized methods to address these challenges will be developed in subsequent sections.   

\subsection{Bayesian inference for the basic random variables}\label{sec: Bayes}
In a structural system, sources of randomness may include variations in material properties, geometric quantities, initial and boundary conditions, dynamic excitation, and environmental effects \cite{beck2002bayesian, ching2007transitional, straub2015bayesian}. A computational model for the system introduces further uncertainty into these sources of randomness. Moreover, the control system, if present, may have additional random quantities. We define $\vect{X} \in \rn$ to represent the basic random variables considered in the Digital Twin system. The term ``basic" indicates that all the other random quantities of the modeled system are deterministic functions of $\vect X$. The distribution of $\vect{X}$ is denoted by $f(\vect{x})$. To ensure computational feasibility, we let the dimension of $\vect{X}$ be finite, implying that random fields and stochastic processes are represented by a finite number of basis functions. Let $\vect{Y} \in \mathbb{R}^{m}$ represent the measured quantities, and $\mathcal{M}: \vect{x} \in\rn\mapsto \vect{y}_{\mathcal{M}} \in \mathbb{R}^{m}$ denote an end-to-end computational model that maps outcomes of $\vect{X}$ into predictions of $\vect{Y}$. The combined effect of modeling and measurement uncertainties is described by $f_{\vect{\epsilon}}(\vect{\epsilon})$, where $\vect{\epsilon} = \vect{y} - \vect{y}_{\mathcal{M}}$ is the difference between the measured $\vect{y}$ and the model prediction $\vect{y}_{\mathcal{M}}$. Note that to construct Bayesian inference for $\vect{X}$, the combined effect of modeling and measurement uncertainties should not be treated as a basic random variable. The posterior distribution of the basic random variables $\vect{X}$, given the knowledge of measurement $\{\vect{Y} = \vect{y}\}$, is:
\begin{equation}
    f(\vect x|\vect y)\propto {f_{\vect{\epsilon}}(\vect{y} - \mathcal{M}(\vect x))}f(\vect x)\,,
\end{equation}
where $f(\vect x)$ and $f(\vect x|\vect y)$ are respectively the prior and posterior distributions, and $f_{\vect{\epsilon}}(\vect{y} - \mathcal{M}(\vect x))$ is the likelihood function.

%\begin{figure}
 % \centering
 % \includegraphics[scale=0.6]{fig3.pdf}
 % \caption{Definition of prior knowledge and current for RT.}
 % \label{Fig:2}
%\end{figure}

In practice, since data is collected in a temporal sequence, the posterior distribution may evolve with time. This process is expressed by the following recursive equation: 
\begin{equation}\label{eq:xtyt}
\begin{aligned}
f(\vect x^{(t)}|\vect y^{(1:t)})\propto&\int_{\rn} f(\vect x^{(t)},\vect x^{(t-1)},\vect y^{(1:t)})\,d\vect x^{(t-1)}\\ 
\propto&\int_{\rn} f(\vect y^{(t)}|\vect x^{(t)},\vect x^{(t-1)},\vect y^{(1:t-1)})f(\vect x^{(t)}|\vect x^{(t-1)},\vect y^{(1:t-1)})f(\vect x^{(t-1)}|\vect y^{(1:t-1)})\,d\vect x^{(t-1)}\\
=&\int_{\rn} f(\vect y^{(t)}|\vect x^{(t)})f(\vect x^{(t)}|\vect x^{(t-1)})f(\vect x^{(t-1)}|\vect y^{(1:t-1)})\,d\vect x^{(t-1)}\\
%\propto&f_{\vect{\epsilon}}(\vect{y}^{(t)} - \mathcal{M}(\vect x^{(t)}))f(\vect x^{(t)}|\vect y^{(1:t-1)})\,,
\end{aligned}
\end{equation}
where subscripts such as $t$ and $1:t$ are introduced to explicitly show the evolving nature of the conditional distributions, $\vect y^{(1:t)}$ denotes data accumulated from a reference time point ``1" to the current $t$, $f(\cdot|\vect y^{(1:t)})$ is the posterior to be established at the current step, and $f(\cdot|\vect y^{(1:t-1)})$ is the posterior from the previous step. The first two lines of Eq.~\eqref{eq:xtyt} are simply marginalization and conditioning rules. The last line assumes a conditional independence structure similar to that in hidden Markov models. 

The transition law for basic random variables, $f(\vect{x}^{(t)}|\vect{x}^{(t-1)})$ in Eq.~\eqref{eq:xtyt}, varies with the nature of $\vect{X}$. If the distribution of $\vect{X}$ remains constant over time, for instance, when $\vect{X}$ represents the elastic moduli or geometries of a structural system, we can assign $f(\vect{x}^{(t)}|\vect{x}^{(t-1)}) = \delta(\vect{x}^{(t)} - \vect{x}^{(t-1)})$. Conversely, if the distribution changes significantly over time, such as when $\vect{X}$ represents the location and magnitude of occasional loads like wind and snow, we can assign $f(\vect{x}^{(t)}|\vect{x}^{(t-1)}) = f(\vect{x}^{(t)})$. These two extreme scenarios for $f(\vect x^{(t)}|\vect x^{(t-1)})$ can simplify Eq.~\eqref{eq:xtyt} into:
\begin{equation}\label{eq:xtyt1}
f(\vect x^{(t)}|\vect y^{(1:t)})\propto\left\lbrace
\begin{aligned}
&f_{\vect{\epsilon}}(\vect{y}^{(t)} - \mathcal{M}(\vect x^{(t)}))f(\vect x^{(t)}|\vect y^{(1:t-1)})\,,&&\text{if }f(\vect{x}^{(t)}|\vect{x}^{(t-1)}) = \delta(\vect{x}^{(t)} - \vect{x}^{(t-1)})\\
&f_{\vect{\epsilon}}(\vect{y}^{(t)} - \mathcal{M}(\vect x^{(t)}))f(\vect x^{(t)})\,,&&\text{if }f(\vect{x}^{(t)}|\vect{x}^{(t-1)}) = f(\vect{x}^{(t)})
\end{aligned}\right.
\end{equation}
Introducing a hyperparameter $\alpha \in [0,1]$ to control the amount of information inherited from the previous posterior, Eq.~\eqref{eq:xtyt1} can be generalized into:
\begin{equation}\label{eq:xtyt2}
f(\vect x^{(t)}|\vect y^{(1:t)})\propto f_{\vect{\epsilon}}(\vect{y}^{(t)} - \mathcal{M}(\vect x^{(t)}))\left(\alpha f(\vect x^{(t)}|\vect y^{(1:t-1)})+(1-\alpha)f(\vect{x}^{(t)})\right)\,.
\end{equation}

Depending on the specifications of $\vect{X}$ and $\vect{Y}$, the inference equation  can predict latent states of a system. The normalizing constant for Eq.~\eqref{eq:xtyt2}, denoted by $c^{(t)}$, can be expressed by: 
\begin{equation}\label{eq:c}
\begin{aligned}
c^{(t)}=&\frac{\alpha^{t-1}}{\prod_{i=1}^{t-1}c^{(i)}}\mathbb{E}_{\vect X}\left[\prod_{i=1}^tf_{\vect\epsilon}(\vect y^{(i)}-\mathcal{M}(\vect X))\right]+(1-\alpha)\sum_{i=2}^{t-1}\frac{\alpha^{t-i}}{\prod_{j=1}^{t-i}c^{(t-j)}}\mathbb{E}_{\vect X}\left[\prod_{j=0}^{t-i}f_{\vect\epsilon}(\vect y^{(t-j)}-\mathcal{M}(\vect X))\right]\\
&+(1-\alpha)\mathbb{E}_{\vect X}\left[f_{\vect\epsilon}(\vect y^{(t)}-\mathcal{M}(\vect X))\right]\,,t>1\,.
\end{aligned}
%\begin{aligned}
%  \prod_{i=1}^tc^{(i)}&=\int_{\rn}f(\vect x)\prod_{i=1}^tf_{\vect{\epsilon}}(\vect{y}^{(i)} - \mathcal{M}(\vect x))\,d\vect x\\
%  &=\mathbb{E}_{\vect X}\left[\prod_{i=1}^tf_{\vect{\epsilon}}\left(\vect{y}^{(i)} - \mathcal{M}(\vect X)\right)\right]\,.
%\end{aligned}  
\end{equation}
Similarly, a generalized moment of interest, $\mathbb{E}_{\vect X|\vect y^{(1:t)}}\left[q(\vect X)\right]$, can be expressed by:
\begin{equation}\label{eq:q}
\begin{aligned}
\mathbb{E}_{\vect X|\vect y^{(1:t)}}\left[q(\vect X)\right]=&\frac{1}{c^{(t)}}\left(\frac{\alpha^{t-1}}{\prod_{i=1}^{t-1}c^{(i)}}\mathbb{E}_{\vect X}\left[q(\vect X)\prod_{i=1}^tf_{\vect\epsilon}(\vect y^{(i)}-\mathcal{M}(\vect X))\right]\right.\\
&+(1-\alpha)\sum_{i=2}^{t-1}\frac{\alpha^{t-i}}{\prod_{j=1}^{t-i}c^{(t-j)}}\mathbb{E}_{\vect X}\left[q(\vect X)\prod_{j=0}^{t-i}f_{\vect\epsilon}(\vect y^{(t-j)}-\mathcal{M}(\vect X))\right]\\
&\left.+(1-\alpha)\mathbb{E}_{\vect X}\left[q(\vect X)f_{\vect\epsilon}(\vect y^{(t)}-\mathcal{M}(\vect X))\right]\right)\,,t>1\,.\\
%\mathbb{E}_{\vect X|\vect y^{(1:t)}}\left[q(\vect X)\right]=\dfrac{\mathbb{E}_{\vect X}\left[q(\vect X)\prod_{i=1}^tf_{\vect{\epsilon}}\left(\vect{y}^{(i)} - \mathcal{M}(\vect X)\right)\right]}{\prod_{i=1}^tc^{(i)}}\,.
\end{aligned}
\end{equation}

\subsection{Risk Shadow}\label{sec:rs}
A Risk Shadow visualizes the reliability indices of structural components. The reliability index provides a unified, comparable metric of risk across various structural components. For a generic component, the reliability index at time step $t$, denoted by $\beta^{(t)}$, is defined as follows:
\begin{equation}\label{eq:beta}
    \beta^{(t)}\defi-\Phi^{-1}(\mathbb{P}_{\vect X|\vect y^{(1:t)}}(\mathcal{R}))\,,
\end{equation}
where $\Phi^{-1}$ is the standard normal inverse cumulative distribution function and $\mathbb{P}_{\vect X|\vect y^{(1:t)}}(\mathcal{R})$ is the probability of a risk event $\mathcal{R}$ measured by $f(\vect x|\vect{y}^{(1:t)})$. Without loss of generality, the event $\mathcal{R}$ is defined as:
\begin{equation}
    \mathcal{R}\defi\{\vect x\in\rn:G(\vect x)\leq0\}\,,
\end{equation}
where the limit-state function $G(\vect x)$ hinges on a computational model that maps $\vect x$ into performance variables to determine whether $\mathcal{R}$ occurs. A typical limit-state function is $G(\vect x)=\text{Capacity}(\vect x)-\text{Demand}(\vect x)$, although more general forms may be required for specialized applications \cite{der2022structural}. 

%\begin{figure}[H]
%  \centering
%  \includegraphics[scale=0.4]{fig2.pdf}
%  \caption{Conceptual Risk Shadow of a wind turbine: physical twin, Digital Twin and Risk Twin.}
%  \label{Fig:3}
%\end{figure}

To compute $\beta^{(t)}$, we express $\mathbb{P}_{\vect X|\vect y^{(1:t)}}(\mathcal{R})$ as:
\begin{equation}\label{eq:pf}
\begin{aligned}
\mathbb{P}_{\vect X|\vect y^{(1:t)}}(\mathcal{R})=&\frac{\Phi(-\beta^{(0)})}{c^{(t)}}\left(\frac{\alpha^{t-1}}{\prod_{i=1}^{t-1}c^{(i)}}\mathbb{E}_{\vect X|\mathcal{R}}\left[\prod_{i=1}^tf_{\vect\epsilon}(\vect y^{(i)}-\mathcal{M}(\vect X))\right]\right.\\
&+(1-\alpha)\sum_{i=2}^{t-1}\frac{\alpha^{t-i}}{\prod_{j=1}^{t-i}c^{(t-j)}}\mathbb{E}_{\vect X|\mathcal{R}}\left[\prod_{j=0}^{t-i}f_{\vect\epsilon}(\vect y^{(t-j)}-\mathcal{M}(\vect X))\right]\\
&\left.+(1-\alpha)\mathbb{E}_{\vect X|\mathcal{R}}\left[f_{\vect\epsilon}(\vect y^{(t)}-\mathcal{M}(\vect X))\right]\right)\,,t>1\,,
%\mathbb{P}_{\vect X|\vect y^{(1:t)}}(\mathcal{R})&=\dfrac{\Phi(-\beta^{(0)})}{\prod_{i=1}^tc^{(i)}}\int_{\rn}f(\vect x |\mathcal{R})\prod_{i=1}^tf_{\vect{\epsilon}}(\vect{y}^{(i)} - \mathcal{M}(\vect x))\,d\vect x\\
%&=\dfrac{\Phi(-\beta^{(0)})}{\prod_{i=1}^tc^{(i)}}\mathbb{E}_{\vect X|\mathcal{R}}\left[\prod_{i=1}^tf_{\vect{\epsilon}}\left(\vect{y}^{(i)} - \mathcal{M}(\vect X)\right)\right]\,,
\end{aligned}
\end{equation}
where $\beta^{(0)}$ is the reliability index associated with the prior distribution of $\vect X$. Eq.~\eqref{eq:pf}, Eq.~\eqref{eq:c}, and Eq.~\eqref{eq:q} are specifically designed to facilitate a ``simulation-free" updating of the Risk Twin. The general idea is that in an offline phase, we can precompute some samples and quantities, so that in the online phase of Risk Twin updating, we only need to plug in observations $\vect y^{(1:t)}$ into simple algebraic equations, resulting in a negligible computational cost. Specifically, in the offline phase, we need to compute $\beta^{(0)}$ using a rare event simulation method, and then prepare the datasets:
\begin{equation}\label{eq:data}
    \begin{aligned}
        &\mathcal{D}=\{(\vect x^{(i)},\mathcal{M}(\vect x^{(i)}))\}_{i=1}^N\,,\vect x^{(i)}\sim f(\vect x)\,,\\
        &\mathcal{D}_{\mathcal{R}}=\{(\vect x^{(i)},\mathcal{M}(\vect x^{(i)}))\}_{i=1}^{N'}\,,\vect x^{(i)}\sim f(\vect x|\mathcal{R})\,.
    \end{aligned}
\end{equation}
Using these prepared datasets and $\beta^{(0)}$, the expectations in Eq.~\eqref{eq:pf}, Eq.~\eqref{eq:c}, and Eq.~\eqref{eq:q} can be readily approximated when observations $\vect y^{(i)}$ are collected, thereby achieving a simulation-free approach in the online phase. The accuracy of this simulation-free approach depends on the number of precomputed samples and the accumulation of time steps. This latter issue can be  mitigated by periodically setting the Risk Twin offline to update $\beta^{(0)}$, $\mathcal{D}$, and $\mathcal{D}_{\mathcal{R}}$ with respect to the recent $f(\vect x|\vect y^{(1:t)})$. It is worth mentioning that the time step in $\vect y^{(1:t)}$ defines a computational mesh that may differ from the real time points for collecting sensor data. For example, one can aggregate sensor data collected at multiple small consecutive time steps into a single $\vect y^{(i)}$ to improve computational efficiency. 

\subsection{Human-Risk Shadow interaction}\label{sec:rc}
We envision risk control being conducted through Human-Risk Shadow interactions. The optimal action $a^{\star}$ at a specific time step depends on the Risk Shadow and the constraints of the control system. We formulate $a^{\star}$ as the solution to the following optimization problem:
\begin{equation}
\begin{aligned}
a^{\star}&=\mathop{\arg\min}_{a\in\Omega_a}\text{Cost}(a,\text{Risk Shadow})\\
&\text{s.t.}\,\left\lbrace\begin{aligned}
&\vect h(a)=\vect 0\\
&\vect g(a)\preceq 0\\
&\tilde{\vect{\beta}}(a)\succeq\vect\beta_0
\end{aligned}\right.
\end{aligned}
\end{equation}
where $\vect h$ and $\vect g$ are equality and inequality constraints from the control system, such as geometry and power constraints, $\tilde{\vect{\beta}}(a)$ denotes the estimated reliability index at the target structural component after an action, $\vect\beta_0$ is the reliability index constraints for the systems. It should be noted that $\tilde{\vect{\beta}}(a)$ can be approximated based on the up-to-date information because the future data is still unavailable.

\section{Benchmark Physical Experiments}\label{sec:be}
This section details the implementation of two benchmark experiments to showcase the capabilities of RT. These illustrative experiments employ materials and equipment that are easily accessible to engineers. The first experiment features a simply supported plate, illustrating the real-time Bayesian inference (Section \ref{sec: Bayes}) for quantifying the uncertainty in position and magnitude of an external force. The second experiment is a cantilever beam controlled by a mechanical arm, showcasing the Risk Shadow (Section \ref{sec:rs}) and human-Risk Shadow interactions (Section \ref{sec:rc}). Video demonstrations of these two experiments are provided below.
\begin{itemize}
    \item \textbf{Bayesian inference for a simply supported plate}: \url{https://youtu.be/vVuwe4H075k}.
    \item \textbf{Risk Shadow for a cantilever beam}: \url{https://youtu.be/MBMhvgd8KKM} (beam) and \url{https://youtu.be/XeRB4-JCY8A} (mechanical arm).
    \item \textbf{Risk Shadow-based control for a cantilever beam}: \url{https://youtu.be/MsjypDyqM40}.
\end{itemize}

\subsection{A simply supported plate}
\begin{figure}[t]
  \centering
  \includegraphics[scale=0.6]{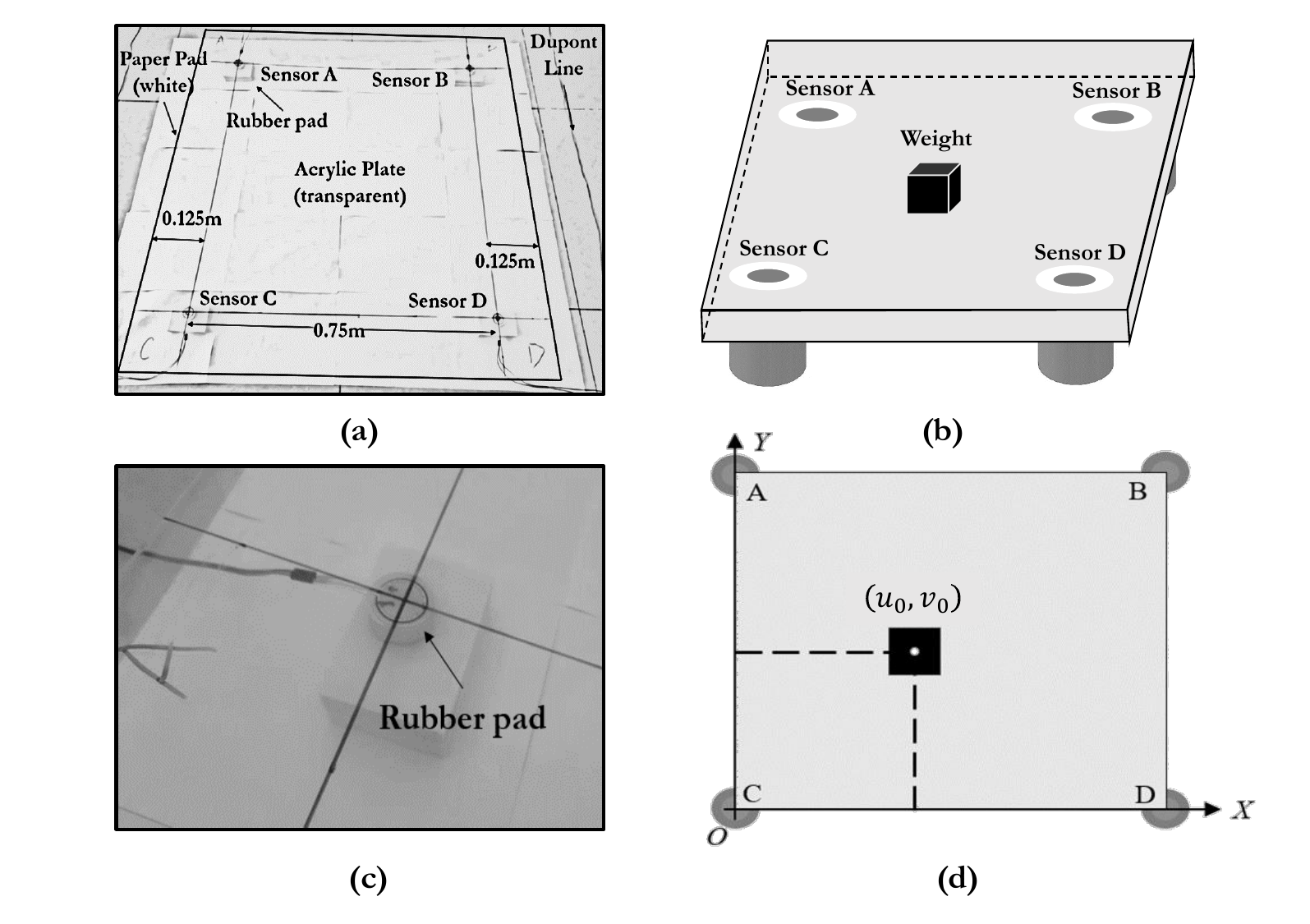}
  \caption{\textbf{A simply supported plate}: (a) the physical system, (b) a sketch illustration, (c) the rubber pad, and  (d) coordinate system for the plate. The physical system consists of a acrylic plate, rubber pads, DuPont wires, sensors, and TTL-USB connectors. The computational platform is developed within the Python environment.}
  \label{Fig:4}
\end{figure}

In this experiment, we constructed a $1\,\text{m}\times1\,\text{m}\times0.05\,\text{m}$ acrylic plate supported by four rubber cushions, each embedded with a stress sensor, as illustrated in Figure \ref{Fig:4}. The sensors, rubber pads, and acrylic plate were affixed using adhesive to prevent slipping. A vertical load was applied to the plate, with the computational goal of quantifying the uncertainty in its magnitude and position.%, as well as analyzing the resulting displacement field.

\subsubsection{Computational model}
Given a weight $w$ located at $(u_0,v_0)$ according to the coordinate system depicted in Figure \ref{Fig:4}(d), the reaction forces at the four supports are:
\begin{equation}
\begin{aligned}
f_A&=\frac{w(l-u_0)v_0}{l^2}\,,\\
f_B&=\frac{wu_0v_0}{l^2}\,,\\
f_C&=\frac{w(l-u_0)(l-v_0)}{l^2}\,,\\
f_D&=\frac{wu_0(l-v_0)}{l^2}\,,\\
\end{aligned}\label{eq:reaction}  
\end{equation}
where $l=0.75$m is the effective side length of the square plate. Therefore, the basic random variables are $\vect X\equiv[W,U_0,V_0]$, the measured quantities are $\vect Y\equiv[F_A,F_B,F_C,F_D]$, and Eq.~\eqref{eq:reaction} defines the computational model $\mathcal{M}: \vect x\in\r^3\mapsto\vect y_{\mathcal M}\in\r^4$ that predicts the measured quantities given an outcome of the basic random variables. 

\subsubsection{Bayesian inference}
\begin{figure}[H]
  \centering
  \includegraphics[scale=0.55]{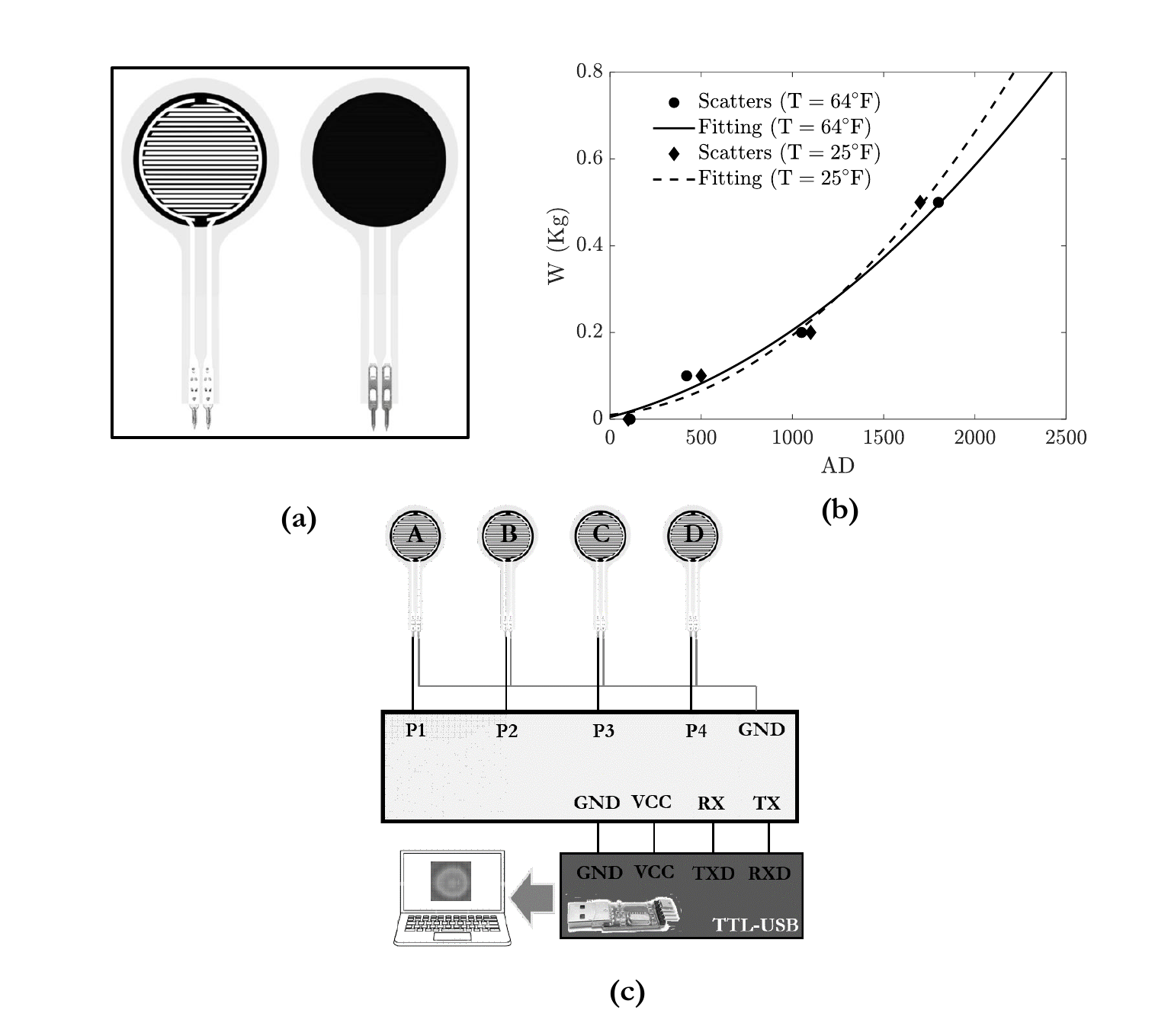}
  \caption{{\textbf{Stress sensors}: (a) flexible thin film pressure sensor with a measuring range of $0-10$ Kg; (b) the force-AD (Analog-Digital) value relationship, measured during daytime and nighttime, where $T$ denotes temperature; (c) the implementation details of the measurement system}.}
  \label{Fig:sensor}
\end{figure}

Uniform priors are assumed for both the location and magnitude of the weight, with the location being anywhere over the plate and the magnitude ranging within $[0, 10]$ kg. It was found that varying temperatures can lead to significant fluctuations in sensor readings, as illustrated in Figure \ref{Fig:sensor}. Additionally, temporal drift also affects measurement accuracy. To address these issues, sensor calibration was conducted using standard weights, and the standard deviation of the sensor measurements was estimated to be $0.1$ kg. The implementation details of the measurement system is illustrated in Figure \ref{Fig:sensor} (c). Assuming a zero-mean Gaussian distribution for the measurement uncertainty, the likelihood function $f_{\vect{\epsilon}}(\vect y-\mathcal{M}(\vect x))$ is:  
\begin{equation}
    f_{\vect{\epsilon}}(\vect y-\mathcal{M}(\vect x))=\exp{\left(-\sum_{i=1}^4\frac{(y_i-\mathcal{M}_i(\vect x))^2}{2\times0.1^2}\right)}\,,
\end{equation}
where $\mathcal{M}_i$ is expressed by the $i$-th line of Eq.~\eqref{eq:reaction}, and $y_i$, $i=1,2,3,4$, are measured forces at supports A, B, C, D, respectively. 

\begin{figure}[H]
  \centering
  \includegraphics[scale=0.55]{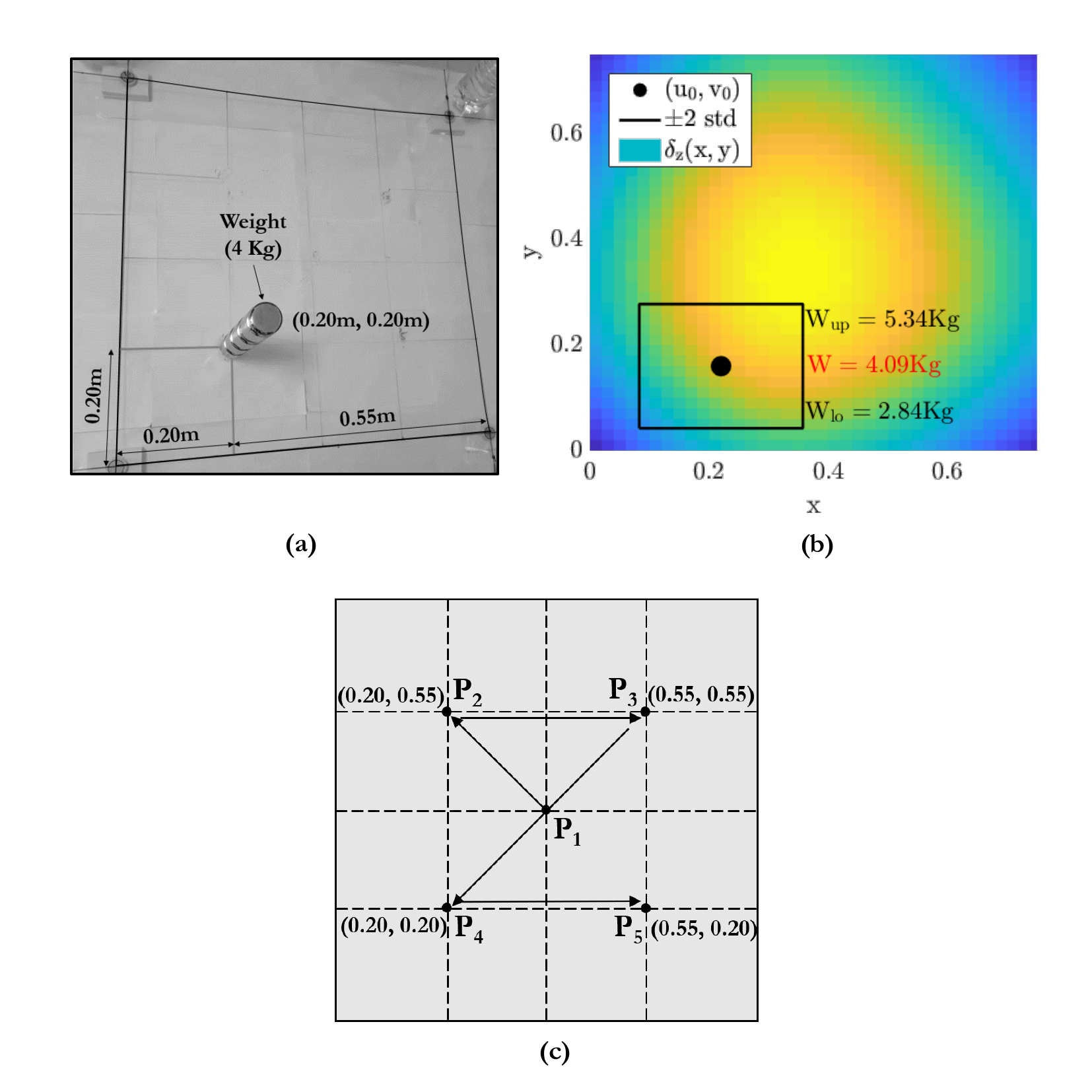}
  \caption{{\textbf{Inference of the position and magnitude of the weight}: (a) the physical system, (b) the inference results for magnitude and location, with two standard deviations of the means, and (c) route for testing real-time inference}. For (c), the weight was moved along the route $P_1\mapsto P_2\mapsto\cdots \mapsto P_5$.  where the video recording can be accessed via \url{https://youtu.be/vVuwe4H075k}.}
  \label{Fig:8}
\end{figure}

To perform inference, we prepared $10^5$ samples during the offline phase, and the $\alpha$ in Eq.~\eqref{eq:xtyt2} is set to $0.5$. Notice that the effect of $\alpha$ is to dilute the previous posterior, used to construct the current prior; the inference results remain insensitive to $\alpha$ as long as $\alpha\neq1$. Using the posterior mean as a point representation, Figure \ref{Fig:8} illustrates an inference result. Subsequently, we shift the weight among five target positions, as illustrated in Figure \ref{Fig:8}(c), and conduct the inference in real-time. The accuracy of these results is documented in Table \ref{table:2}, and a video demonstrating the process is available at \url{https://youtu.be/vVuwe4H075k}.

\begin{table}[H]
\centering
\caption{Accuracy of Bayesian inference for $P_1$, $P_2$, $P_3$, $P_4$ and $P_5$.}
\begin{tabular}{c c c c c} 
 \toprule
Position & $\Delta d$ & $\Delta d/l$ & $\Delta W$ & $\Delta W/W$ \\ [0.5ex] \midrule
 $P_1$ & 0.018 & 2.40\% & 0.01 & 0.25\% \\ 
 $P_2$ & 0.052 & 6.93\% & 0.11 & 2.75\% \\
 $P_3$ & 0.113 & 15.07\% & 0.32 & 8.00\% \\
 $P_4$ & 0.087 & 11.60\% & 0.23 & 5.75\% \\
 $P_5$ & 0.023 & 3.07\% & 0.74 & 18.5\% \\[1ex] 
 \bottomrule
\end{tabular}
\label{table:2}
\end{table}

\subsection{A cantilever beam controlled by a mechanical arm}
\begin{figure}[H]
  \centering
  \includegraphics[scale=0.55]{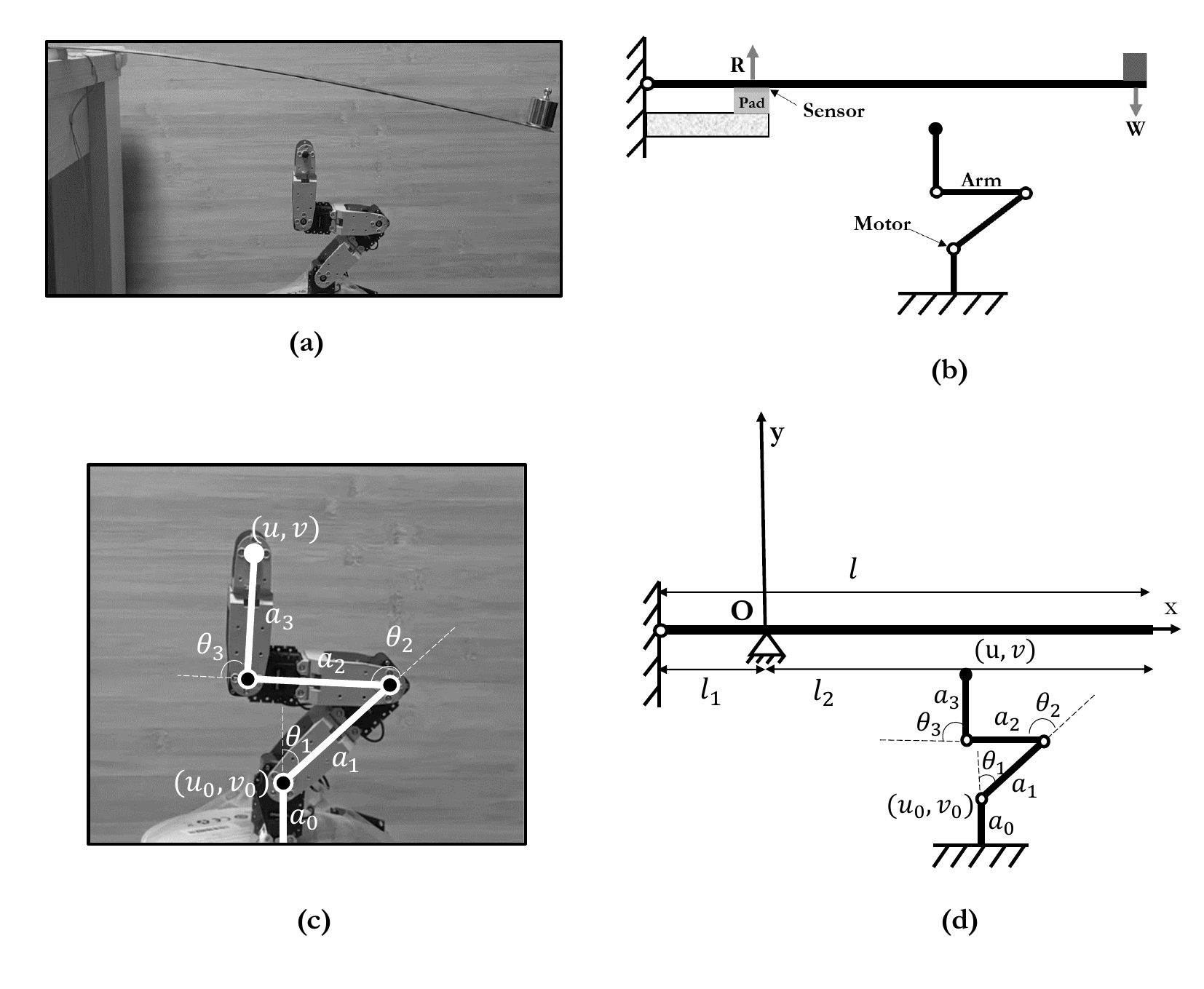}
  \caption{\textbf{A cantilever beam controlled by a mechanical arm}: (a) the physical system, (b) a sketch illustration, (c) the schemes of mechanical arm with physically dimensional sizes, and (d) definition of plane coordinates.}
  \label{Fig:10}
\end{figure}

In this experiment, we constructed a cantilever beam with dimensions $1\,\text{m} \times 0.04\,\text{m} \times 0.002\,\text{m}$, controlled by a mechanical arm, as illustrated in Figure~\ref{Fig:10}. The left end of the beam is fixed in vertical and horizontal directions, and a vertical support, equipped with a pressure sensor identical to those used in the previous experiment, is located $0.4\,\text{m}$ from the left end. The base of the mechanical arm is positioned at $(0.3\,\text{m}, -0.3\,\text{m})$, in accordance with the coordinate system depicted in Figure~\ref{Fig:10}(d). The mechanical arm comprises three motors and high-strength, lightweight structural components. A weight is applied at the right end of the beam, and the mechanical arm is utilized to mitigate the deflection.

\subsubsection{Computational model}\label{section:3.2.1}
The geometric and mechanical properties of the beam and the mechanical arm are summarized in Table \ref{table:3}. The basic random variables considered in this experiment include the weight of the load, $W$, the maximum allowable stress of the beam, $\mathrm{\Sigma}_{\max}$, the maximum allowable torque of the mechanical arm, $M_{\max}$, the inaccuracies in motor rotations, $\Delta\Theta_i$, where $i=1,2,3$, and the vertical control force of the mechanical arm, $F_c$. The measured quantity is the reaction force at the support, denoted by $R$. Therefore, the basic random variables are $\vect X\equiv[W,\mathrm{\Sigma}_{\max},M_{\max},\Delta\Theta_1,\Delta\Theta_2,\Delta\Theta_3,F_c]$, and the measured quantity is $Y\equiv R$. The model $\mathcal{M}$ that predicts the measured quantity given an outcome of the basic random variables is:
\begin{equation}
    \mathcal{M}(\vect x) = \left\lbrace\begin{aligned}
    &\frac{l}{l_1}w\,,&&\text{uncontrolled phase}\,,\\
    &\frac{l}{l_1}w-\frac{l_c+l_1}{l_1}f_c\,,&&\text{controlled phase}\,,
    \end{aligned}\right.
\end{equation}
where $l_c$ is the horizontal location of the control force, expressed by
\begin{equation}
   l_c = u_0+\sum_{i=1}^3\left(a_i\sin(\sum_{j=1}^i(\theta_j+\Delta\theta_j))\right)\,,
\end{equation}
where $u_0$ is the horizontal position of the base of the mechanical arm, $a_i$ is the length of the $i$-th arm component, and $\theta_i$ is the rotational angle of the $i$-th motor subjected to uncertainty specified by $\Delta\Theta_i$. Clockwise rotation is considered as the positive direction and counterclockwise is the negative. Notice that $\theta_i$ is known from the control system of the mechanical arm. 

Similar to the previous experiment, the likelihood function is: 
\begin{equation}
    f_{\vect{\epsilon}}(y-\mathcal{M}(\vect x))=\exp{\left(-\frac{(y-\mathcal{M}(\vect x))^2}{2\times0.1^2}\right)}\,.
\end{equation}
The limit-state function for the beam at any location $u\in(0,l_2)$ is defined as:
\begin{equation}
    G_b(\vect x;u)=\left\lbrace\begin{aligned}
   & \sigma_{\max}-\frac{w(l_2-u)h_b}{2I}\,,&&\text{uncontrolled phase or $u>l_c$ controlled phase}\,,\\
   &\sigma_{\max}-\frac{w(l_2-u)h_b}{2I}+\frac{f_c(l_c-u)h_b}{2I}\,,&&\text{$u<l_c$, controlled phase}\,,
    \end{aligned}\right.
\end{equation}
which indicates failure when the stress at any location $u$ exceeds $\sigma_{\max}$. The limit-state functions for each motor of the mechanical arm are defined as:
\begin{equation}\label{eq:Ga}
    G_a^i(\vect x)=m_{\max}-m_i\,,
\end{equation}
where $m_i$ denotes the moment at the $i$-th motor, expressed by:
\begin{equation}
\begin{aligned}
  m_1=\left\lbrace\begin{aligned}
   & G_m\left(a_1\sin(\theta_1+\Delta\theta_1)+\sum_{j=1}^2\left(a_j\sin(\sum_{k=1}^j(\theta_k+\Delta\theta_k))\right)\right)\equiv m_{G1}\,,&&\text{uncontrolled phase}\,,\\
&m_{G1}+f_c\sum_{j=1}^3\left(a_j\sin(\sum_{k=1}^j(\theta_k+\Delta\theta_k))\right)\,,&&\text{controlled phase}\,,
    \end{aligned}\right.     
\end{aligned}\label{eq:m1}
\end{equation}
\begin{equation}
    m_2=\left\lbrace\begin{aligned}
   & G_m\left(a_2\sin(\sum_{j=1}^2(\theta_j+\Delta\theta_j))\right)\equiv m_{G2}\,,&&\text{uncontrolled phase}\,,\\
&m_{G2}+f_c\sum_{j=2}^3\left(a_j\sin(\sum_{k=1}^j(\theta_k+\Delta\theta_k))\right)\,,&&\text{controlled phase}\,,
    \end{aligned}\right.\label{eq:m2}
\end{equation}
\begin{equation}
    m_3=\left\lbrace\begin{aligned}
   & 0\,,&&\text{uncontrolled phase}\,,\\
&f_c\left(a_3\sin(\sum_{j=1}^3(\theta_k+\Delta\theta_k))\right)\,,&&\text{controlled phase}\,,
    \end{aligned}\right.\label{eq:m3}
\end{equation}
where $G_m$ is the weight of the motor.  

\begin{table}[H]
\centering
%\caption{Parameters for the beam and mechanical arm 2, where $\theta_m$ denotes the output mean value of degrees from the ports linking to three motors and $\theta_{orig}$ denotes the reposition angle of the three motors. A clockwise direction constitutes a positive rotation, while a counterclockwise direction denotes a negative rotation.}
\caption{Parameters for the beam and mechanical arm.}
\begin{tabular}{c c c c c} 
 \toprule
Component & Parameters & Type & Mean & Standard Deviation \\ [0.5ex] \midrule
 & $l$ & Deterministic & 1.0 m & - \\ 
 & $l_1$ & Deterministic & 0.4 m & - \\
 & $l_2$ & Deterministic & 0.6 m & - \\
 & $h_b$ & Deterministic & 0.002 m & - \\
  Beam & $b_b$ & Deterministic & 0.04 m & - \\
 & I & Deterministic & $2.67\times10^{-11}\,\text{m}^4$ & - \\
 & $E$ & Deterministic & 200GPa & - \\
 & $\mathrm{\Sigma}_{\max}$ & Lognormal & 250Mpa & 25MPa \\
% & $\varphi_s$ & Normal & 0 & 0.1Kg \\ [0.5ex] \hline
\midrule
 & $a_0$ & Deterministic & 0.052 m &- \\ 
 & $a_1$ & Deterministic & 0.093 m & - \\
 & $a_2$ & Deterministic & 0.093 m & - \\
 & $a_3$ & Deterministic & 0.078 m & - \\
 Arm & $G_m$ & Deterministic & 5.5 N & - \\
 & $\Delta\Theta_i$, $i=1,2,3$ & Normal & $0^{\circ}$ & $3^{\circ}$ \\
% & $\vect{\theta}_{orig}$ & Deterministic & $[\pi/4,-3\pi/4,\pi/2]$ & N/A \\
% & $\vect{\theta}_{thr}$ & Deterministic & $[\pi/4,-3\pi/4,\pi/2]$ & N/A \\
 & $M_{\max}$ & Lognormal & $1.5\,N\cdot m$ & $0.15\,N\cdot m$ \\
 & $(u_0,v_0)$ & Deterministic & (0.3,-0.3) m & - \\[1ex] 
\bottomrule
\end{tabular}
\label{table:3}
\end{table}

\subsubsection{Risk Shadow}\label{section:3.2.3}
To construct the Risk Shadow, we prepared $10^5$ samples for $\mathcal{D}$ and $10^3$ samples for $\mathcal{D}_\mathcal{R}$ (recall Eq.~\eqref{eq:data}) during the offline phase, and the $\alpha$ in Eq.~\eqref{eq:xtyt2} is set to $0.5$. This sample size can be insufficient for highly reliable components. Therefore, if a probability is estimated to be zero through Eq.~\eqref{eq:pf}, we apply the first-moment second-order method to obtain a preliminary estimate for visualization. The first-moment second-order analysis has been adapted to incorporate the offline and online phases for rapid inference. Sensor data are transmitted to the computational platform at a rate of $10$ times per second, thus the Risk Shadow has an image refresh frequency of $10$ Hz. We visualize reliability indices in the Risk Shadow using the color map in Table \ref{table:4}. The color map was chosen as a preliminary illustration of the Risk Shadow; it may require further investigation regarding scalability and differentiability. 

\begin{figure}[H]
  \centering
  \includegraphics[scale=0.68]{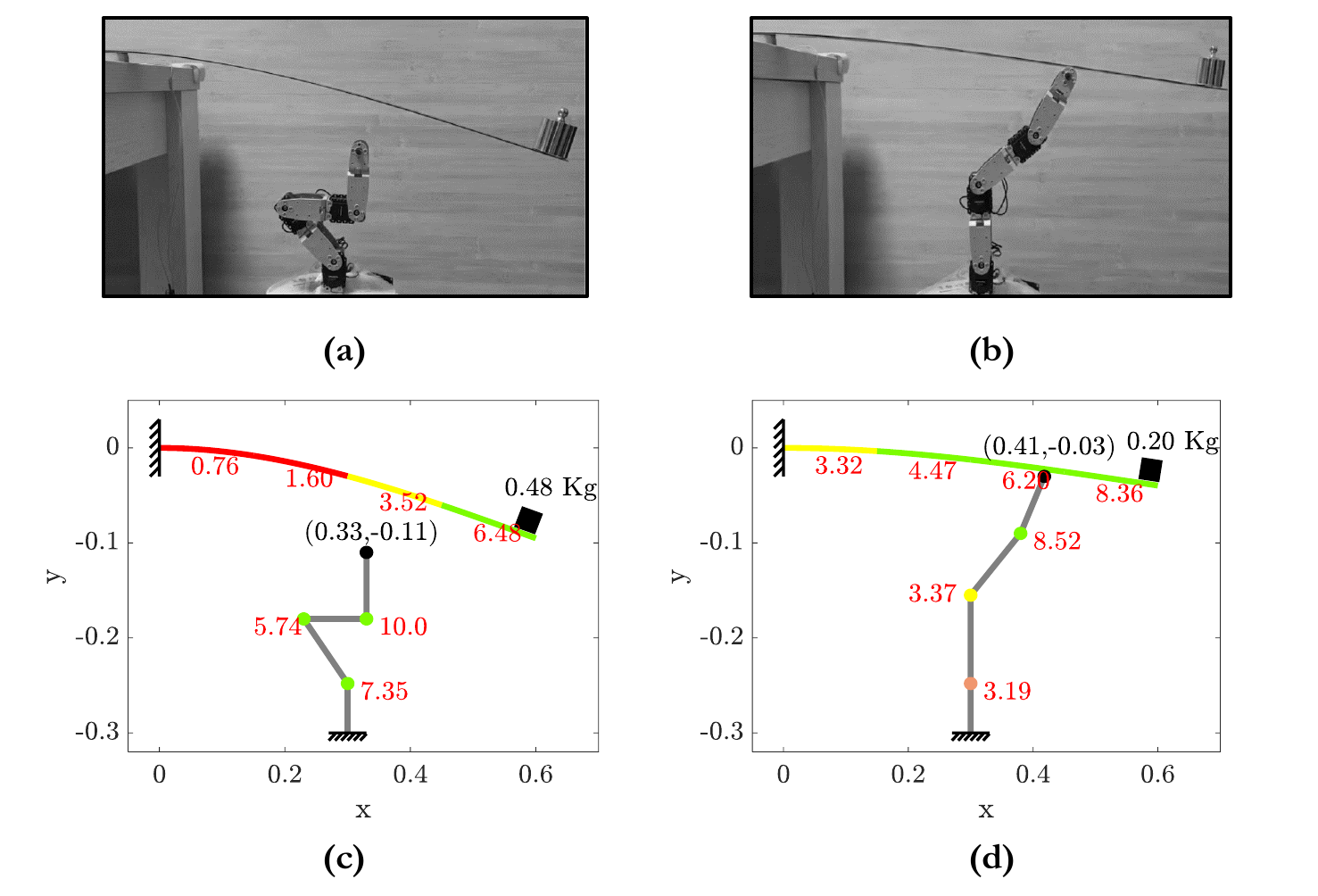}
  \caption{\textbf{Snapshots of the Risk Twin with}: (a) physical system in the uncontrolled phase, (b) physical system in the controlled phase, (c) Risk Shadow of (a) and (d) Risk Shadow of (b). In (a), a 0.5 kg weight is placed on the right end, and the system is in the uncontrolled phase. In (b), a 0.2 kg weight is placed on the right end, while the system is under the control of the mechanical arm. In (c) and (d), the beam is divided into four segments, displaying the maximum reliability indices within each segment. Furthermore, the reliability indices of the three motors are also displayed.}
  \label{Fig:15}
\end{figure}

Snapshots of the Risk Shadow are illustrated by Figure \ref{Fig:15}. It is worth mentioning that the third motor does not bear any load when it operates in the free motion state, so we assigned a reliability index of ``10" to suggest the upper bound for the Risk Shadow outputs. More demonstrations of this Risk Twin system can be found in videos at \url{https://youtu.be/MBMhvgd8KKM} and \url{https://youtu.be/XeRB4-JCY8A}.

\begin{table}[H]
\centering
\caption{Color map for Risk Shadow.}
\begin{tabular}{c c c c} 
\toprule
Level of Risk & Reliability index &  Color & RGB \\ [0.5ex] \midrule
Safe & $\beta\geq3.7$ & \includegraphics[scale=1]{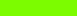}
& [124, 252, 0] \\ 
Low Risk & $3.2\leq\beta<3.7$ & \includegraphics[scale=1]{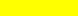} & [255, 255, 0] \\
Medium Risk & $2.7\leq\beta<3.2$ & \includegraphics[scale=1]{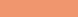} & [240, 150, 110] \\
High Risk & $\beta<2.7$ & \includegraphics[scale=1]{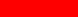} & [255, 0, 0] \\[1ex] 
\bottomrule
\end{tabular}
\label{table:4}
\end{table}

%As illustrated in Fig \ref{Fig:16}, the weight on the right-end changes from 0.1Kg, 0.2Kg, to 0.5Kg adaptively. Moreover, the Risk Shadow of arm is also investigated, where endpoint of arm, $(x_3,y_3)$, moves from (0.4m, -0.2m) to (0.4m, -0.1m), (0.2m, -0.1m) and (0.2m, -0.2m), adaptively. It reveals that while there may be a delay of approximately 0.1s in updating the physical twin's state, the RT achieves synchronization and frequency consistency, rendering it a feasible and effective tool for reflecting the structural system's risk status. Although we cannot display the RT 's video dynamic effect in text format, we have uploaded the video demonstration online for interested readers to explore. Demonstrations of Risk Shadow are uploaded on the website: (i) \url{https://youtu.be/MBMhvgd8KKM} for the bending beam and (ii) \url{https://youtu.be/XeRB4-JCY8A} for the mechanical arm. Moreover, this demonstration presents instructive potential for application in more complex level for systems or structures.

%\begin{figure}
%  \centering
%  \includegraphics[scale=0.68]{fig16.pdf}
%  \caption{Actions of adding loads and rotating arm, where weights changes from 0.1Kg to 0.2Kg and 0.5Kg and the endpoint of mechanical arm changes from (0.4m, -0.2m) to (0.4m, -0.1m), (0.2m, -0.1m) and (0.2m, -0.2m), adaptively.}
%  \label{Fig:16}
%\end{figure}

\subsubsection{Human-Risk Shadow interaction}\label{section:3.2.4}
To implement Risk Shadow-based control for this experiment, we propose cost-benefit control actions that satisfy geometrical and reliability constraints. Specifically, the 
geometrical constraint is 
\begin{equation}
\begin{aligned}
   u - u_0&= \sum_{i=1}^3\left(a_i\sin(\sum_{j=1}^i(\theta_j+\Delta\theta_j))\right)\,,\\
   v - v_0&= \sum_{i=1}^3\left(a_i\cos(\sum_{j=1}^i(\theta_j+\Delta\theta_j))\right)\,,\\
\end{aligned}\label{eq:geo}
\end{equation}
where $(u,v)$ is a generic target endpoint position of the mechanical arm and $\theta_i$ can be controlled. 

The reliability constraint ensures that, after an increment $\delta\theta_i$ for $\theta_i$, the reliability index for each motor does not exceed a threshold $\beta_0$, i.e.,
\begin{equation}\label{eq:reli}
\beta(G_a^i(\vect x;\delta\vect\theta))-\beta_0 \geq 0,, i=1,2,3,
\end{equation}
where $G_a^i(\vect x;\delta\vect\theta)$ denotes the limit-state Eq.~\eqref{eq:Ga} with $\vect\theta$ replaced by $\vect\theta+\delta\vect\theta$.

For a small increment $\delta\vect\theta$, the expected energy cost can be approximated by:
\begin{equation}
C(\delta\vect\theta)\approx\E{\sum_{i=1}^{3}\lvert M_i\delta\theta_i\rvert}\approx\sum_{i=1}^{3}\lvert m_i(\Delta\theta=\vect0)\delta\theta_i\rvert\,, 
\label{eq:cost}
\end{equation}
where the expectation is evaluated using the first-order approximation by fixing the random variables $\Delta\vect{\Theta}$ to their mean values (zeros). Notice that the expressions for $m_i$, recall Eqs.~\eqref{eq:m1}-\eqref{eq:m3}, vary with the uncontrolled (free-motion) and controlled phases. Therefore, it is ideal to adopt a small $\delta\vect{\theta}$ for each control step. Specifically, the action to achieve a target endpoint position $(u_c,v_c)$ of the mechanical arm is discretized into a sequence of small increments $\delta\vect\theta^{(\tau)}$, such that at each step, the energy described by Eq.~\eqref{eq:cost} is minimized while satisfying the constraints given by  Eq.~\eqref{eq:geo} and Eq.~\eqref{eq:reli}. The corresponding optimization problem is defined as: 
\begin{equation}
\begin{aligned}\label{eq:optc}
 \delta\vect\theta^{(\tau)}&=\mathop{\arg\min}_{\delta\vect\theta}C(\delta\vect\theta)\,,\\
 &\text{s.t. }\left\lbrace\begin{aligned}
 &\text{Eq.~\eqref{eq:geo} with $(u,v)=(u^{(\tau)},v^{(\tau)})$}\,,\\
 &\text{Eq.~\eqref{eq:reli}}\,,\\
 \end{aligned}\right.
\end{aligned}
\end{equation}
where the target locations $(u^{(\tau)},v^{(\tau)})$ for the step $\tau$ is set to: 
\begin{equation}
    (u^{\tau},v^{(\tau)})=(u^{0},v^{(0)})+\tau\frac{(u_c,v_c)-(u^{(0)},v^{(0)})}{n_{\tau}}\,,\tau=1,2,...,n_{\tau}\,,
\end{equation}
where a linear interpolation between the target $(u_c,v_c)$ and initial $(u^{0},v^{(0)})$ locations is adopted. This simple linear interpolation, chosen for its simplicity, may not yield a globally minimized energy cost from $(u^{(0)},v^{(0)})$ to $(u_c,v_c)$. \ref{Append:implementationdetails} shows the implementation details for controlling the endpoint of the mechanical arm to a target position. 

\begin{figure}[H]
  \centering
  \includegraphics[scale=0.85]{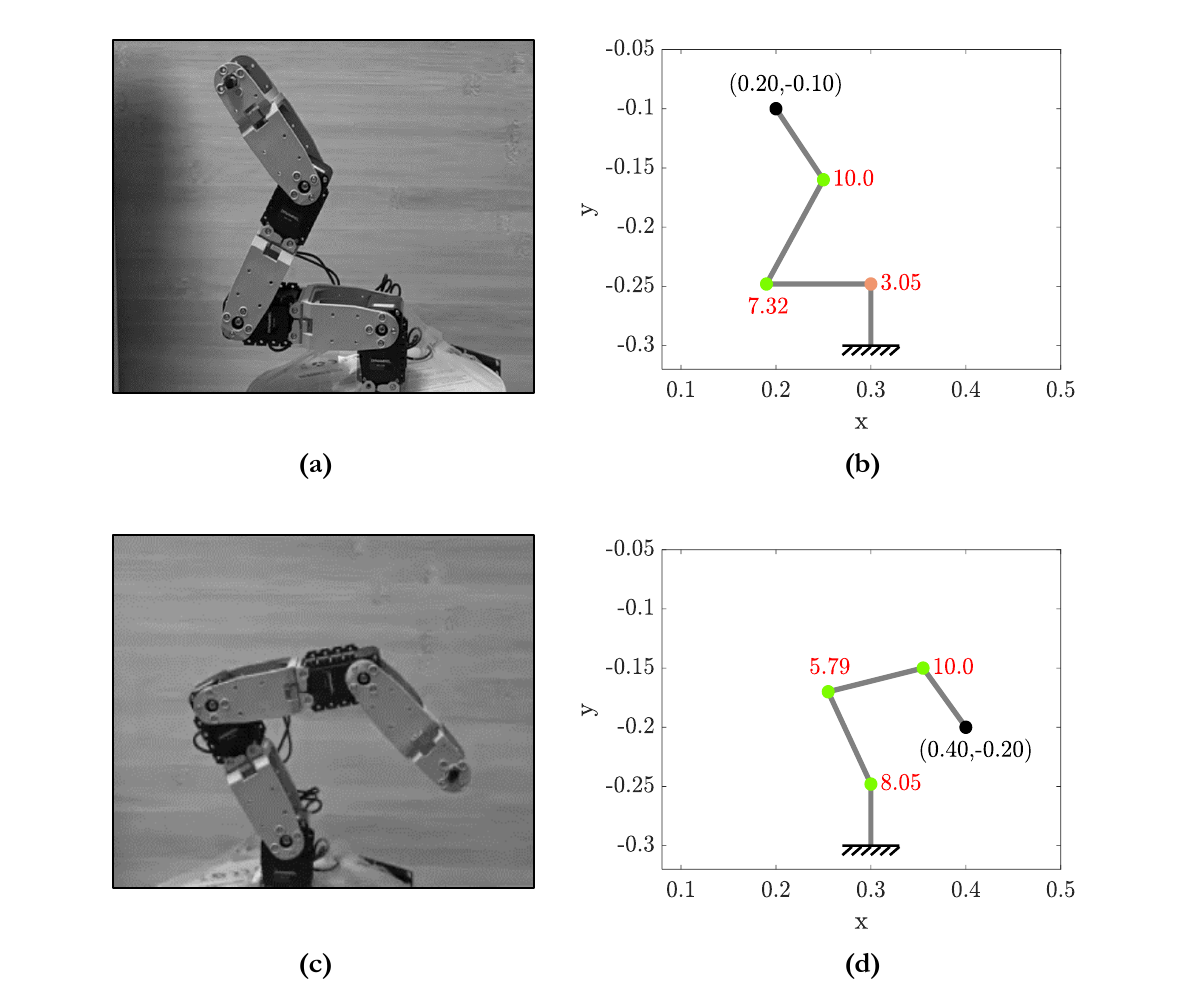}
  \caption{\textbf{Illustration of controlling the mechanical arm}: (a) move the endpoint to the location $(0.20, -0.10)$; (b) Risk Shadow of (a); (c)  move the endpoint to the location $(0.40, -0.20)$; (d) Risk Shadow of (c).}
  \label{Fig:17}
\end{figure}

\begin{figure}[H]
  \centering
  \includegraphics[scale=0.8]{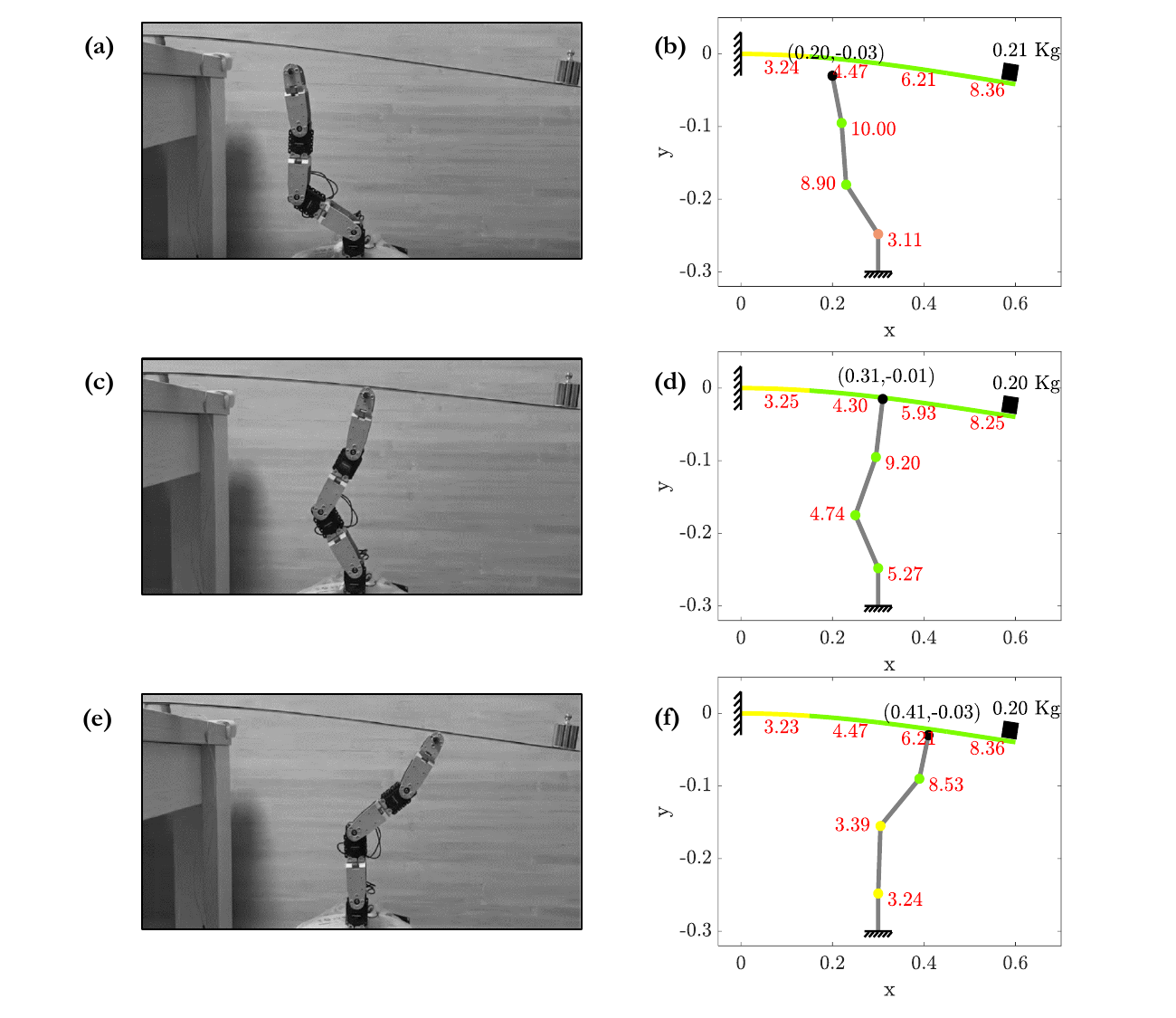}
  \caption{\textbf{Human-Risk Shadow interaction}: (a) control location $u_c = 0.2$ m; (b) Risk Shadow of (a); (c) control location $u_c = 0.3$ m; (d) Risk Shadow of (c); (e) control location $u_c = 0.4$ m; (f) Risk Shadow of (e). To illustrate the interaction between human risk and shadow, we set a hypothetical goal to control the reliability indices of the beam so that they exceed $4.2$, while the reliability indices of the motors should be no less than $3.1$. The first attempt to control the beam at location $u_c=0.20$ m is shown in panels (a) and (b). As the mechanical arm was elevated, the reliability of the motor near the base decreased to $3.11$, indicating that continuing the control would be risky. Consequently, the user adjusted the control location to $u_c=0.30$ m and successfully completed the control task by further elevating the arm. The actual control location achieved was $u_c=0.31$ m, with an error of $1$ centimeter. Panels (e) and (f) illustrate another control attempt at $u_c=0.40$ m. In this scenario, the control action was mechanically more effective compared to the $u_c=0.30$ m scenario, but the reliability indices of the motors were lower.}
  \label{Fig:18}
\end{figure}

Figure \ref{Fig:17} displays the Risk Shadow of the mechanical arm in free motion, with more details illustrated by the demonstration video at \url{https://youtu.be/XeRB4-JCY8A}. Figure \ref{Fig:18} illustrates a control process by human-Risk Shadow interaction, with more details at \url{https://youtu.be/MsjypDyqM40}. 

\section{Risk Twin of a Wind Turbine}\label{sec:wt}
\subsection{Computational model of a wind turbine}
\begin{figure}[H]
  \centering
  \includegraphics[scale=0.9]{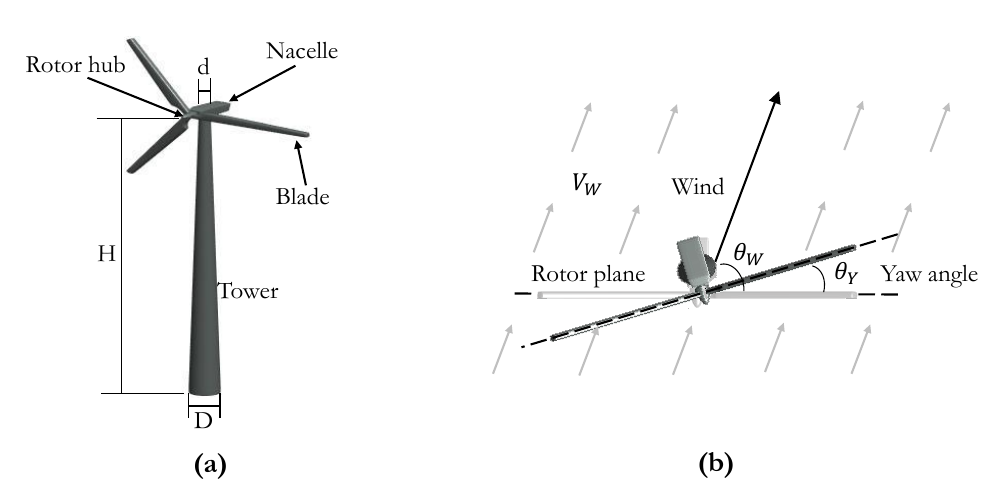}
  \caption{\textbf{A wind turbine structure subject to wind loads}: (a) Conceptual illustration of the four core components in a wind turbine and (b) the angular definitions of yaw angle, $\theta_Y$, wind speed, $V_W$, and wind direction, $\theta_W$. This paper considers only the deviation between $\theta_Y$ and $\theta_W$, without addressing the effects of upwind or downwind.}
  \label{WTFig:9}
\end{figure}

We demonstrate the application of Risk Twin using a simulation model of a wind turbine, where parameters and computational methodologies are partially referred to  \cite{hansen2015aerodynamics}. The proposed RT framework updates and visualizes the reliability indexes of the structural components under three limit states in real-time. Moreover, the uncertainty parameters primarily originate from wind speed, wind direction, sensor equipment, and the load-bearing capacity of the components. As shown in Figure \ref{WTFig:9}, a horizontal-axis wind turbine (HAWT) consists of four key components: a tubular tower, blades, rotor hub, and nacelle. The total height of the tubular tower, $H$, is $80$ meters, including the height of the hub and nacelle. The base diameter of the tower, $D$, is $10$ meters, tapering linearly to an upper diameter $d$ of $4$ meters, with a wall thickness of $0.02$ meters. As documented in \cite{hansen2015aerodynamics}, the variation of the local chord of the blade along the rotor radius is presented in Table \ref{table:4}. The wind turbine is equipped with three blades, each with a mass of $4000$ kg, assuming the center of mass is at the midpoint of the blade. The angular acceleration due to the friction between hub and blade, is inversely proportional to the angular velocity with the coefficient $k_f=0.01$. To prevent the blades from rotating excessively fast and inducing damage, a brake system is installed at the rotor hub. This system activates when the blade's rotational speed reaches a maximum value of $\omega_{max}=1.8$ rad/sec. Once the brake system is activated, the angular acceleration of the blade drops to zero, preventing further increases in angular velocity. We treat the simulated wind turbine as a physical twin, serving as the basis to demonstrate the application of Risk Twin in wind farms. A demonstration video of this simulation model can be accessed through the link: \url{https://youtu.be/2iTJ_FDZXrg}.

\begin{figure}[h]
  \centering
  \includegraphics[scale=.8]{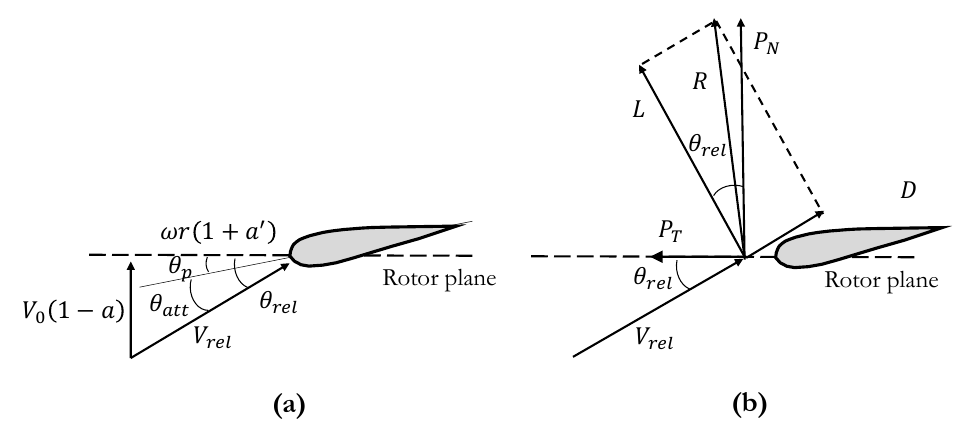}
  \caption{\textbf{Aerodynamics of wind turbine blades}: (a) Relative velocity analysis considering vortex effects and (b) Structural analysis of lift and drag forces per unit length.}
  \label{WTFig:10}
\end{figure}

\begin{table}[H]
\centering
\caption{Geometry of the wind turbine blade: local chord length $c$ and the corresponding radius $r$.}
\begin{tabular}{c c c c c c c c c c } 
 \toprule
 $r$ [m] & 4.5 &  6.5 &  8.5 &  10.5&  12.5 & 14.5 &  16.5&  18.5 &  20.3 \\ [0.5ex] \midrule
 $c$ [m] & 1.63 &  1.540 &  1.420 &  1.294 &  1.163 &  1.026 &  0.881 &  0.705 & 0.265 \\ [1ex] 
 \bottomrule
\end{tabular}
\label{table:4}
\end{table}
Wind turbines operate by harnessing wind force to drive the rotation of the blades, thereby generating mechanical energy in the rotor. This mechanical energy is then converted into electrical energy by the generator located in the nacelle. Therefore, the aerodynamic properties that facilitate the generation of mechanical energy are crucial. Figure \ref{WTFig:10}(a) illustrates the mechanical analysis model of a blade with a pitch angle under operational conditions, subjected to wind speed $V_0$ perpendicular to the rotor plane, and the rationale for computing the relative velocity $V_{rel}$, where $V_0 = V_W\sin{\left| {\theta_{w}-\theta_{Y}} \right|}$. $V_{rel}$ is decomposed into the perpendicular components $(1-a)V_0$ and $(1+a')\omega r$:
\begin{equation}
    V_{rel}\sin{\theta_{rel}}=V_0(1-a)\,,
\end{equation}
and
\begin{equation}
V_{rel}\cos{\theta_{rel}}=V_0(1+a')\,,
\end{equation}
where $\theta_{rel}$ denotes the angle between the rotor plane and $V_{rel}$, $a$ denotes the axial induction factor due to the existing vortex system, $a'$ is the tangential induction factor, $r$ denotes the radial length of the target position on the blade, and $\omega$ is the rotational speed of the rotor. It follows that $V_{rel}$ can be computed as:
\begin{equation}
    V_{rel}=\sqrt{(V_0 (1-a))^2+(\omega r(1+a'))^2}\,,
\end{equation}
and
\begin{equation}
    \tan{\theta_{rel}} = \frac{(1-a)V_0}{(1+a')\omega r}\,.
\end{equation}

\begin{figure}[h]
  \centering
  \includegraphics[scale=0.5]{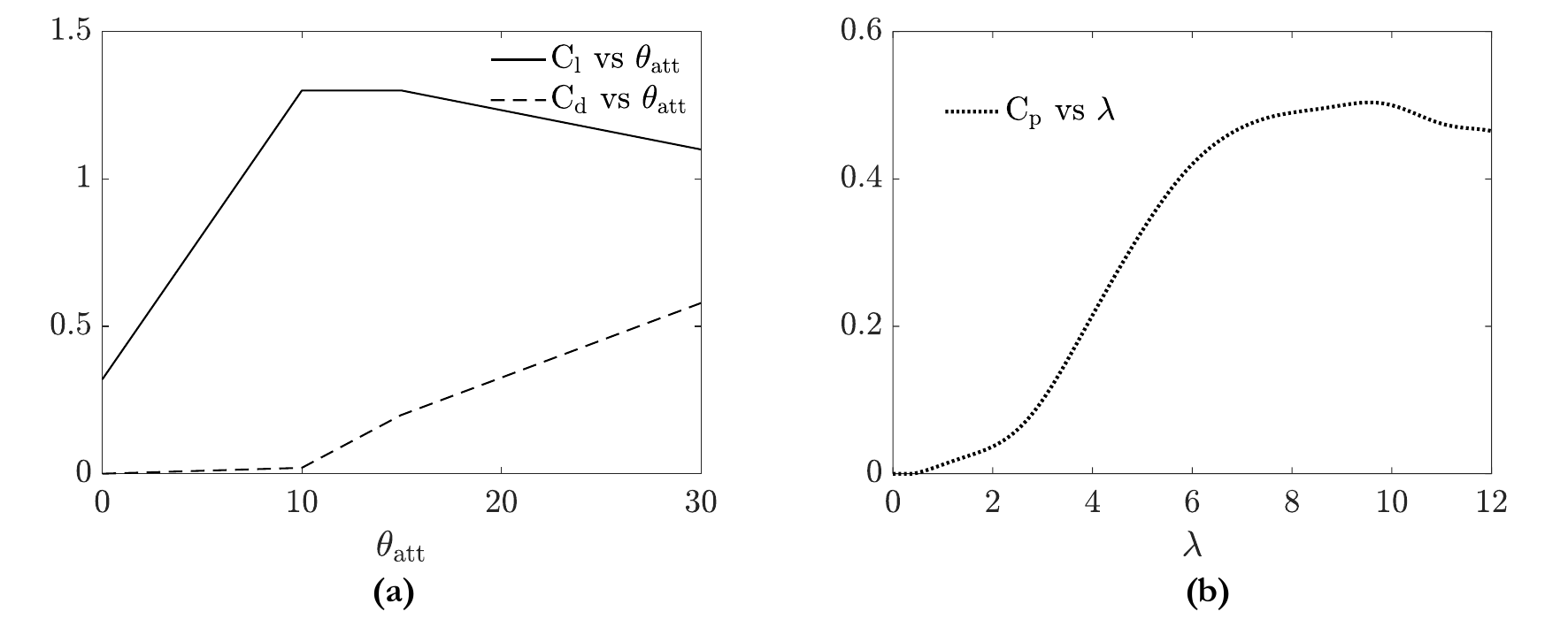}
\caption{\textbf{Relationships between model parameters}: (a) $C_l$ and $C_d$ versus $\theta_{att}$ and (b) $C_p$ versus $\lambda$.}
  \label{WTFig:11}
\end{figure}

The effect of the vortex system can be negligible, indicating that $a = a' = 0$. Moreover, let $\theta_{att} = \theta_{rel} - \theta_p$ be the local angle of attack of the wind. According to the aerodynamics of wind turbines, the lift $L$ and drag force $D$ per unit length caused by the wind can be respectively calculated by:
\begin{equation}
    L=\frac{1}{2}\rho V_{rel}^2 c C_l\,,
\end{equation}
\begin{equation}
    D=\frac{1}{2}\rho V_{rel}^2 c C_d\,,
\end{equation}
where $\rho$ denotes the air density, with $\rho = 1.29 \, \text{kg/m}^3$, and $c$ denotes the local chord of the blade. The coefficients $C_l$ and $C_d$ are the lift and drag coefficients, which are parameterized by $\theta_{att}$. The relationship between $C_l$, $C_d$, and $\theta_{att}$ is illustrated in Figure \ref{WTFig:11}(a). It should be noted that $C_d$ can be significantly large if $\theta_{att} < 0$. However, the more critical forces are those perpendicular to the rotor plane, denoted by $P_N$, and those along the rotor plane direction, denoted by $P_T$. $P_N$ acts along the blade, making the rotor highly susceptible to flapwise moments, potentially causing damage at the blade-hub connection. Additionally, $P_T$ drives the rotor blade's rotation, serving as the primary source of power for the wind turbine. However, excessive $P_T$ can generate an overly large shaft moment, which can lead to destructive bending moments in the hub and nacelle structure. Figure \ref{WTFig:10}(b) illustrates the geometric relationships among $V_{rel}$, $L$, $D$, $P_N$, and $P_T$, expressed as:
\begin{equation}
    P_N = L\cos{\theta_{rel}} + D\sin{\theta_{rel}}\,,
\end{equation}
\begin{equation}
    P_T = L\sin{\theta_{rel}} - D\cos{\theta_{rel}}\,.
\end{equation}
Consequently, the power of the wind turbine can be computed by:
\begin{equation}
    P=\frac{1}{2}\rho {V_0}^3AC_p(\lambda)\,,
\end{equation}
where $A = \pi R^2$ is the rotor area, $\lambda = \omega R / V_0$ denotes the tip-speed ratio, and $C_p(\lambda)$ represents the power coefficient parameterized by $\lambda$. Figure \ref{WTFig:11}(b) illustrates the relationship between $C_p$ and $\lambda$. The coefficients $a$ and $a'$ can be estimated through the classical Blade Element Momentum (BEM) method, expressed as:
\begin{equation}\label{eq:a}
    a=\frac{1}{\frac{4{\sin}^2{\theta_{rel}}}{\sigma C_n}+1}\,,
\end{equation}
and
\begin{equation}\label{eq:ap}
    a'=\frac{1}{\frac{4\sin{\theta_{rel}}\cos{\theta_{rel}}}{\sigma C_t}-1}\,,
\end{equation}
where $\sigma$ denotes the fraction of the annular area in the control volume in BEM, expressed as:
\begin{equation}
    \sigma=\frac{cB}{2\pi r}
\end{equation}
where $B = 3$ is the number of blades. $C_n$ and $C_t$ in Eqs.~\eqref{eq:a} and \eqref{eq:ap} can be estimated by:
\begin{equation}
    C_n = C_l\cos{\theta_{rel}} + C_d\sin{\theta_{rel}}\,,
\end{equation}
and
\begin{equation}
    C_t = C_l\sin{\theta_{rel}} - C_d\cos{\theta_{rel}}\,.
\end{equation}

\begin{figure}[H]
  \centering
  \includegraphics[scale=0.7]{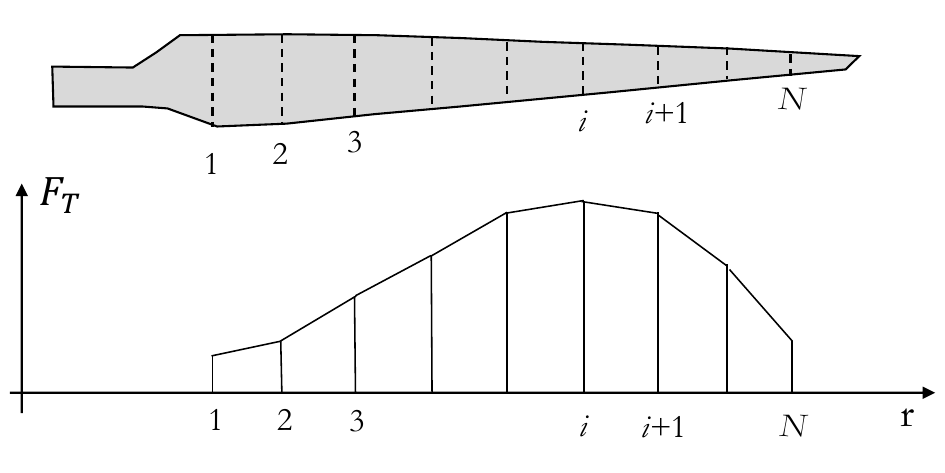}
  \caption{\textbf{Illustration of discretization of elements in blade through BEM method}.}
  \label{WTFig:12}
\end{figure}

Generally, the final values of $a$ and $a'$ are obtained through an iterative process. First, one initializes $a$ and $a'$ and compute $\theta_{rel}$ and $\theta_{att}$ to obtain the initial values of $C_l$ and $C_d$. Next, $C_n$ and $C_t$ are computed to update $a$ and $a'$ until the results converge. The final values of $a$ and $a'$ can be corrected according to correction factors such as Prandtl’s tip loss factor and Glauert correction. As shown in Figure \ref{WTFig:12}, the BEM discretizes the blade into $N$ elements, where $F_T$ denotes the tangential forces for each element. Moreover, it should be noted that the trend of force in normal direction $F_N$ also aligns with $F_T$. $F_T$ and $F_N$ can be computed based on $P_T$ and $P_N$ through the integration of all the elements along blade. Consequently, the shaft torque, $M_{shaft}$, can be calculated via the BEM method:
\begin{equation}
   M_{shaft} = \sum_{i=1}^{N-1}\left(\frac{p_{T,i+1}-p_{T,i}}{3(r_{i+1}-r_i)}(r_{i+1}^3-r_i^3)+\frac{p_{T,i}r_{i+1}-p_{T,i+1}r_i}{2(r_{i+1}-r_i)}(r_{i+1}^2-r_i^2 ) \right)\,,\\
\end{equation}
where $r_i$ denotes the radial length of the $i$-th element, and $p_{T,i}$ is the force per unit length tangential to the rotor plane of the $i$-th element. The rotational speed of the rotor can be computed by:
\begin{equation}
    \dot{\omega} = \frac{M_{shaft}}{B I_b} - k_f \omega
\end{equation}
where $\omega$ and $\dot{\omega}$ represent the rotor's angular speed and acceleration, respectively, and $I_b$ denotes the blade's moment of inertia about the hub. 

\subsection{Structural failures of the wind turbine}
\begin{figure}[h]
  \centering
  \includegraphics[scale=0.85]{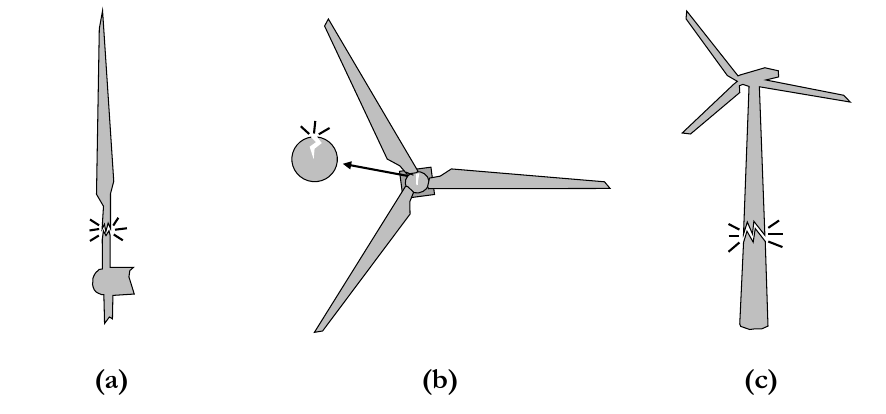}
\caption{\textbf{Conceptual illustration of the three failure modes of the wind turbine}: (a) blade failure due to flapwise moment, (b) hub failure due to shaft moment, and (c) tower failure due to stress.}
  \label{WTFig:13}
\end{figure}

The wind turbine has a probability of encountering extreme wind loads during operation, which poses a risk of structural failure. We consider the failure modes of the blade, rotor hub, and tower. As shown in Figure \ref{WTFig:13}, blade failure primarily arises from the flapwise direction moment, hub failure is mainly due to excessive shaft moment from the brake system, and tower failure occurs when the maximum stress exceeds the threshold value. Therefore, we define three limit state functions.

The limit-state function for blade failure is defined as the flapwise bending moment exceeding a threshold, expressed as \cite{hu2012reliability}:
\begin{equation}\label{WTlsf:1}
    g_{blade}=M_{flap}^{thr}-\left| {M_{flap}} \right|\,,
\end{equation}
where
\begin{equation}
   M_{flap} = \sum_{i=1}^{N-1}\left(\frac{p_{N,i+1}-p_{N,i}}{3(r_{i+1}-r_i)}(r_{i+1}^3-r_i^3)+\frac{p_{N,i}r_{i+1}-p_{N,i+1}r_i}{2(r_{i+1}-r_i)}(r_{i+1}^2-r_i^2 ) \right)\,.
\end{equation}

The limit-state function for the rotor hub failure is defined as the shaft moment exceeding a threshold is defined as follows:
\begin{equation}\label{WTlsf:2}
    g_{hub}=M_{shaft}^{thr}-\left| {M_{shaft}} \right|\,,
\end{equation}
where
\begin{equation}
    M_{shaft}=M_{shaft}-BIk_{f}{\omega}\,.
\end{equation}

The tower is modeled as a bending beam subjected to a point load $F_N$ at the top and a distributed wind pressure that increases with height along the tower. The limit state function is defined as the maximum stress induced by these forces exceeding a threshold, expressed as:
\begin{equation}\label{WTlsf:3}
    g_{tower}=\sigma_{tower}^{thr}-\max_h \sigma_{tower}\,,
\end{equation}
and
\begin{equation}
    \sigma_{tower}(h)=\sqrt{\sigma_1^2(h)+\sigma_2^2(h)}\,,
\end{equation}
where $\sigma_{tower}(h)$ denotes the stress of the tower at the height $h$, $\sigma_1$ denotes the stress caused by normal force $F_N$ and $\sigma_2$ is the stress induced by wind pressure. $\sigma_1$ and $\sigma_2$ can be computed as:
\begin{equation}
    \sigma_i(h)=\frac{M_i(h)D_h(h)}{2I_t},\ i=1,2\,,
\end{equation}
where $D_h$ denotes the diameter of the tower at $h$, $I_t$ is the moment of inertia of the cross-section, $M_1$ denotes the equivalent moment caused by the normal force $F_N$ expressed by:
\begin{equation}
    M_1(h)=F_N(H-h)\,,
\end{equation}
and $M_2$ can be computed by:
\begin{equation}
    M_2(h)=\frac{1}{2}\rho\int_{h}^{H} (z-h)V_h^2(z)\eta D_h(z) \, dz\,,
\end{equation}
where $\eta$ is the gust factor that considers the ratio between maxima and mean of the wind velocity, and the wind velocity at the height $h$, $V_h(z)$, can be computed as:
\begin{equation}
    V_h(z)=V_w(z/H)^{0.15}\,.
\end{equation}

\subsection{Risk Twin of the wind turbine}
\begin{figure}[h]
  \centering
  \includegraphics[scale=0.6]{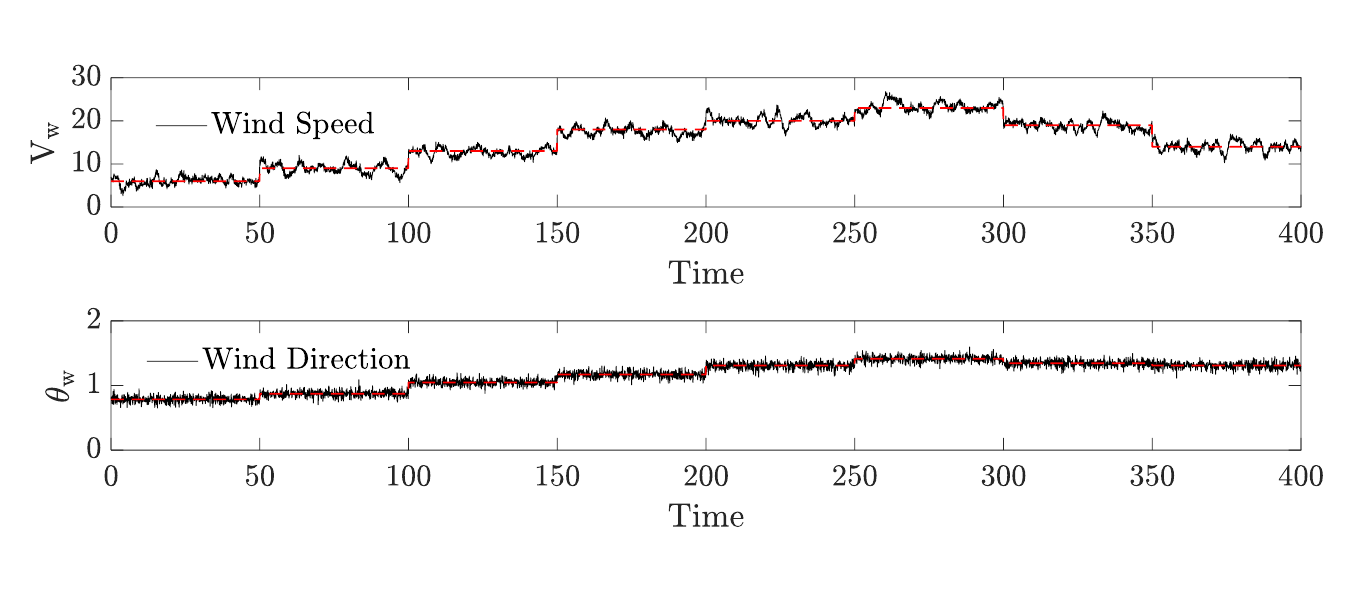}
\caption{\textbf{A random realization of the wind speed and wind direction.}}
  \label{WTFig:14}
\end{figure}

We consider the uncertainties arising from the thresholds of the limit state functions, wind speed and direction, and their measurements, as summarized in Table \ref{table:5}. The wind speed $V_w$ and wind direction $\theta_w$ are modeled by homogeneous Gaussian processes. Specifically, the wind speed is modeled by the following equation:
\begin{equation}
    V_w(t)=\Psi(t,\kappa_w),
\end{equation}
where $t$ denotes the operation time of wind turbine and $\Psi$ denotes a Gaussian process. The wind direction, $\theta_w$, is modeled by another Gaussian process expressed by:
\begin{equation}
    \theta_w(t)=\Theta(t,\kappa_w ),
\end{equation}
where $\Theta$ is a Gaussian process. The kernel/correlation functions of both Gaussian processes are modeled by:
\begin{equation}
    \kappa_w(t_1,t_2)=\sigma_{f_{s,d}}^2\exp{\left(\frac{-(t_1-t_2)^2}{2\tau}\right)}\,,
\end{equation}
where $\sigma_{f_{s,d}}^2$ denotes the process variance of wind speed or direction with $\sigma_{f_{s}}^2=1\ m/s$ and $\sigma_{f_{d}}^2=1^{\circ}$, and $\tau$ is the correlation length scale of the process. In this paper, we set $\sigma_f = 1$ and $\tau = 1$. Figure \ref{WTFig:14} shows a random realization of the wind speed and direction processes, where the black solid line is the random realization and red dashed line represents mean value. We assume sensors to measure the wind speed and direction are installed at the nacelle. The measurement errors $\phi_v$ and $\phi_{\theta}$ for wind speed and direction, respectively, follow Gaussian distributions with parameters listed in Table \ref{table:5}. 

\begin{table}[h]
\centering
\caption{Random variables of wind turbine.}
\begin{tabular}{c c c c} 
 \toprule
Random Variable & Distribution Type & Mean & Standard Deviation \\ [0.5ex] \midrule
 $M_{flap}^{thr}$ & Lognormal & $2\times10^5(N\cdot m)$ & $1\times10^4(N\cdot m)$ \\ 
 $M_{shaft}^{thr}$ & Lognormal & $2\times10^5(N\cdot m)$ & $1\times10^4(N\cdot m)$ \\
 $\sigma_{tower}^{thr}$ & Lognormal & $7.5\times10^6(N/m^2)$ & $7.5\times10^5(N/m^2)$ \\
 $V_w$ & Gaussian Process & Time-Variant & 1.5m/s \\
 $\theta_w$ & Gaussian Process & Time-Variant & $3^{\circ}$ \\
 $\phi_v$ & Normal & 0 & 0.5 m/s \\ 
 $\phi_{\theta}$ & Normal & 0 & $3^{\circ}$ \\ [1ex] 
 \bottomrule
\end{tabular}
\label{table:5}
\end{table}

We conduct two simulation experiments: (i) the \textit{forward experiment}, which shows the information flow from the simulated physical model to the digital model, where the data collected from sensors are used to build the Risk Shadow, and (ii) the \textit{inverse experiment}, which shows the information flow from the digital model to the physical model, where optimization of the yaw angle $\theta_Y$ and pitch angle $\theta_p$ is performed. For both experiments, the total simulation time is $400$ seconds, with initial parameters $\theta_Y = 0^{\circ}$ and $\theta_p = 5^{\circ}$. In the inverse experiment, $\theta_Y$ and $\theta_p \in [0, 5^{\circ}]$ are adaptively tuned, with specific steps detailed in Algorithm \ref{alg:wt control} in \ref{Append:implementationdetails}.

\begin{figure}
    \centering
    \begin{subfigure}{0.75\textwidth}
        \centering
        \includegraphics[scale=0.75]{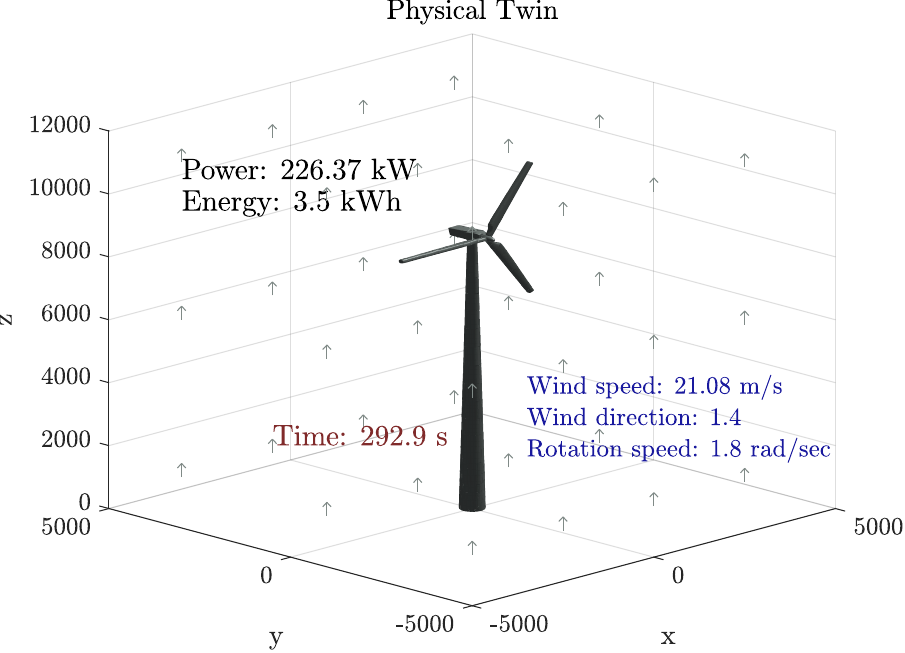}
        \caption{}
    \end{subfigure}
    \hfill
    \begin{subfigure}{0.75\textwidth}
        \centering
        \includegraphics[scale=0.75]{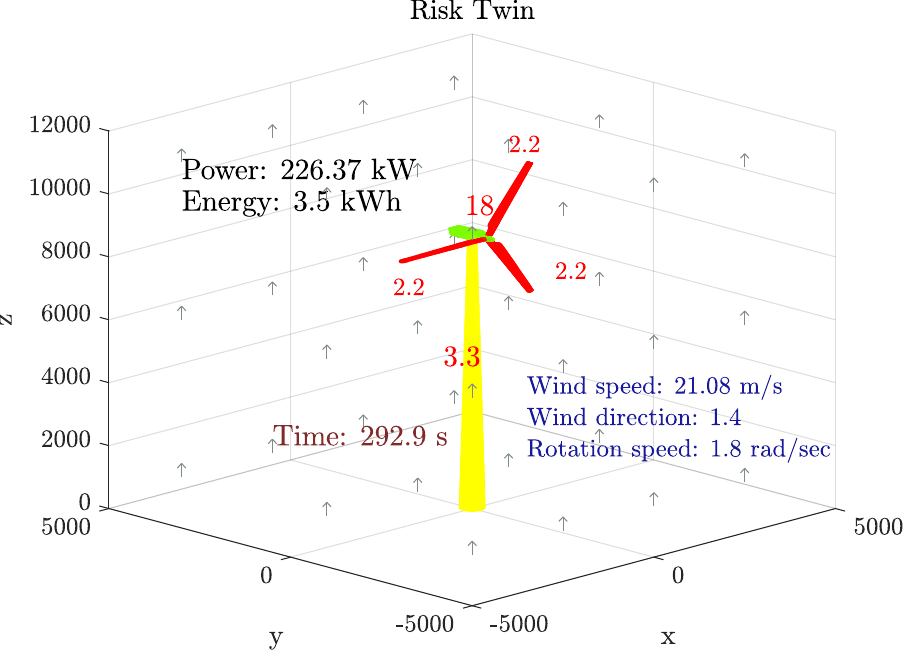}
        \caption{}
    \end{subfigure}
    \caption{\textbf{Risk Twin of simulated wind turbine}: (a) simulated physical twin and (b) Risk Twin.}
    \label{WTFig:15}
\end{figure}

Figure \ref{WTFig:15} shows snapshots at $292.9$ seconds from the forward experiment. Figure \ref{WTFig:15}(a) depicts the physical twin of the simulated wind turbine, while the corresponding Risk Twin is shown in Figure \ref{WTFig:15}(b). The wind speed at this moment is $V_w = 21.08 \, \text{m/s}$, approaching gale force, and the wind direction is $\theta_w = 1.4 \, \text{rad}$. Therefore, the wind turbine is in a high-risk condition. According to Figure \ref{WTFig:15}(b), the hub is in a safe state with $\beta_{hub} = 18$, the blade is in a high-risk state with $\beta_{blade} = 2.2$, and the tower is in a low-risk state with $\beta_{tower} = 3.3$. The Risk Shadow effectively visualizes the structural risks.

\begin{figure}
    \centering
    \begin{subfigure}{0.75\textwidth}
        \centering
        \includegraphics[scale=0.75]{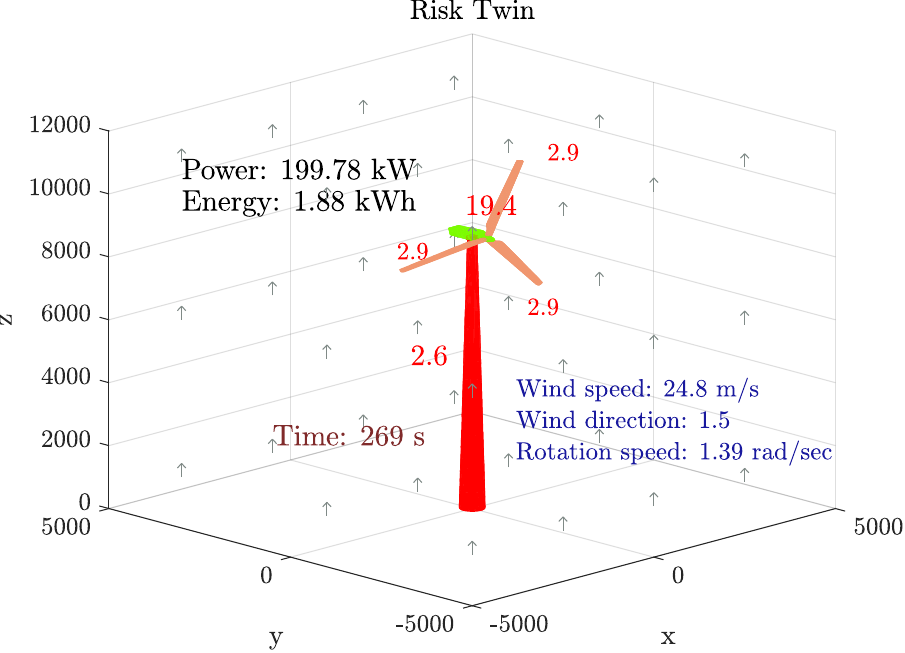}
        \caption{}
    \end{subfigure}
    \hfill
    \begin{subfigure}{0.75\textwidth}
        \centering
        \includegraphics[scale=0.75]{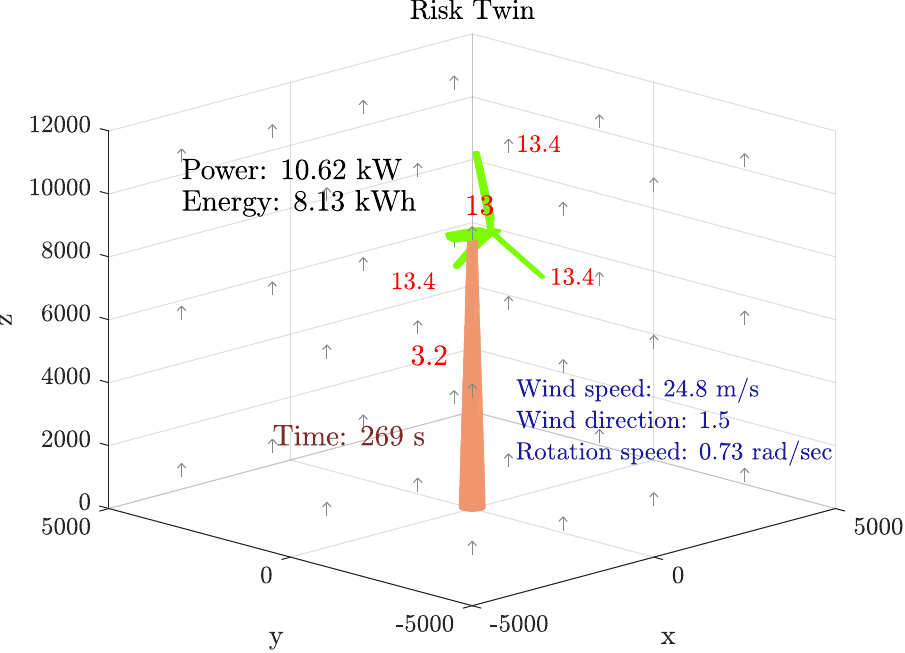}
        \caption{}
    \end{subfigure}
    \caption{\textbf{Risk Twin of simulated wind turbine}:  (a) fixed parameters and (b) optimized parameters.}
    \label{WTFig:16}
\end{figure}

Figure \ref{WTFig:16} shows snapshots at $269$ seconds from the inverse experiment. Figure \ref{WTFig:16}(a) depicts the Risk Twin based on fixed parameters, while Figure \ref{WTFig:16}(b) represents the one with optimized parameters. It is seen that the optimization effectively mitigates the potential risks to the wind turbine under intense wind load conditions. Specifically, $\theta_Y$ was optimized from $0^\circ$ to $70.5^\circ$, and $\theta_p$ was optimized from $5^\circ$ to $29.5^\circ$. Additionally, the reliability index $\beta_{blade}$ was increased from $2.9$ to $13.4$, and $\beta_{tower}$ was increased from $2.6$ to $3.2$. The figure also shows a trade-off between power generation and structural reliability. 

\begin{figure}[h]
  \centering
  \includegraphics[scale=0.75]{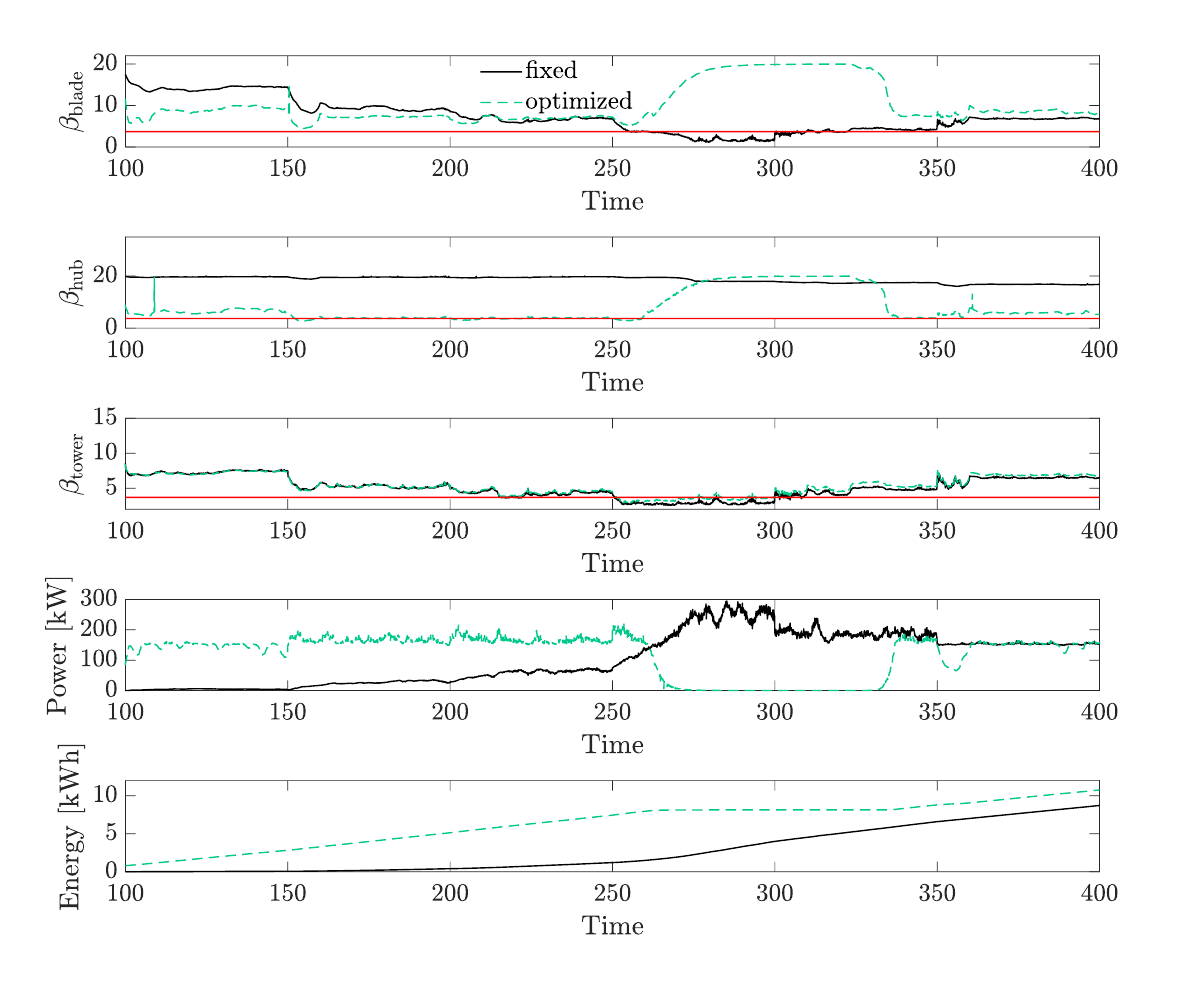}
\caption{\textbf{Performance of wind turbine with optimization}: (a) posterior reliability index of the blade, (b) posterior reliability index of the hub, (c) posterior reliability index of the tower, (d) real-time power, and (e) produced energy. The red solid line denotes the reliability threshold. During the period from 250 seconds to 350 seconds, when wind speeds are high, the control system of the Risk Twin seeks to reduce the structural risk of the blade, hub, and tower. During the period from 50 seconds to 250 seconds, when wind speeds are low, the control system seeks to increase wind power generation.}
  \label{WTFig:17}
\end{figure}

Figure \ref{WTFig:17} compares the reliability indices and energy generation of the wind turbine with and without dynamic yaw and pitch angle controls, i.e., the inverse flow from the digital model to the physical model. The black solid line represents the simulation with fixed yaw and pitch angles, while the blue dashed line represents the results with dynamic angle controls. Several observations can be made: (i) The proposed Risk Twin can enhance the power generation efficiency of the wind turbine during low wind speeds. (ii) The Risk Twin can help the structure avoid potential failures during high wind speeds. (iii) It seeks a balance between structural safety and power generation. 

\section{Conclusions}\label{Sec:conclude}
This paper introduces and demonstrates the concept of Risk Twin, a Digital Twin system designed for real-time reliability visualization and reliability-informed control. The computational challenges of statistical inference and reliability updating are addressed by decomposing the service life of Risk Twins into offline and online phases. During the offline phase, computationally intensive models are simulated to prepare datasets and quantities for rapid, simulation-free inference in the online phase. To illustrate the proposed concept, two benchmark physical experiments and one simulation experiment are conducted. The first physical experiment demonstrates real-time Bayesian inference for quantifying the uncertainty in the position and magnitude of an external force on a simply supported plate. The second physical experiment showcases the Risk Shadow and Human-Risk Shadow interactions using a cantilever beam controlled by a mechanical arm. The final numerical experiment showcases the bi-directional information flow between physical and digital models using a simulated wind turbine. Future research could explore the application and adaptation of Risk Twin technology to real-world civil structures and infrastructures.

\section*{Acknowledgement}
\noindent
The first author is partly supported by the New Faculties' Basic Research Capability Enhancement Program through No. 2-9-2022-010 at China University of Geosciences (Beijing). The first author appreciates the assistance of Yi Ren at Qualcomm (Shanghai) for his support in sensor connectivity and Zhuoqun Gao for her meticulous care. The authors appreciate the constructive comments from Prof. Daniel Straub at the Technical University of Munich. Any opinions, findings, and conclusions expressed in this paper are those of the authors and do not necessarily reflect the views of the sponsors.

\appendix
\section{Implementation details of the mechanical arm and wind turbine control}\label{Append:implementationdetails}

\begin{algorithm}[H]
\caption{Reliability-informed control of the mechanical arm}\label{alg:arm control}
\begin{algorithmic}[1]
    \State \textbf{a.} Read parameters, $\theta_i$, from the mechanical arm with the current position;
    \State \textbf{b.} Compute the current position $\vect{p} = (u^{0}, v^{0})$ of the endpoint of the mechanical arm;
    \State \textbf{c.} Define the number of discretizations $n_{\tau}$;
    \State \textbf{d.} Define $\vect{p}_c = (u_c, v_c)$.
    \State Compute $\Delta{\vect{p}} = (\vect{p}_c - \vect{p}) / n_{\tau}$;
    \If {the mechanical arm touches the beam}
        \State \textbf{a.} Yes, compute $m_i$ using the controlled phase equation;
        \State \textbf{b.} No, compute $m_i$ using the uncontrolled phase equation.
    \EndIf
    \State $\vect{p} \leftarrow \vect{p} + \Delta{\vect{p}}$ and estimate $\delta{\vect{\theta}}^{\star}$ based on Eq.~\eqref{eq:optc};
    \State Control the arm moving to $\vect{p}$ using $\delta{\vect{\theta}}^{\star}$;
    \If {$\vect{p}$ reaches $\vect{p}_c$}
        \State \textbf{a.} Yes, stop;
        \State \textbf{b.} No, go back to step 3.
    \EndIf
\end{algorithmic}
\end{algorithm}

\begin{algorithm}[H]
\caption{Reliability-informed control of the wind turbine}\label{alg:wt control}
\begin{algorithmic}[1]
    \State \textbf{a.} Read parameters from Risk Twin;
    \State \textbf{b.} Define the control parameters $\Delta\theta$ and $\beta_{thr}$.
    \State Define the proposed decisions, $D_i$:
    \State \hspace{1em} $\theta_Y(D_i) \leftarrow \theta_Y + m \cdot \Delta\theta$, 
    \State \hspace{1em} $\theta_p(D_i) \leftarrow \theta_p + n \cdot \Delta\theta$, where $m,\ n \in [-1, 0, 1]$;
    \State Estimate the prior reliability $\hat{\beta}(D_i)$ and power $\hat{P}(D_i)$ of the $D_i$:
    \State \hspace{1em} \textbf{a.} $\hat{\beta}(D_i) = \min{[\beta_{blade}(D_i), \beta_{hub}(D_i), \hat{\beta}_{tower}(D_i)]}$.
    \State \hspace{1em} \textbf{b.} $\hat{P}(D_i) = Power[D_i]$.
    \If{$\hat{\beta}(D_i) \geq \beta_{thr}$}
        \State \textbf{a.} Yes, select decision $D_i$ that maximizes $\hat{P}(D_i)$, go back to step 1;
    \Else
        \State \textbf{b.} No, select decision $D_i$ that maximizes $\hat{\beta}(D_i)$, go back to step 1.
    \EndIf
\end{algorithmic}
\end{algorithm}

\bibliography{Main}
	
\end{document}